\pdfoutput=1

\documentclass[conference]{IEEEtran}

\IEEEoverridecommandlockouts

\usepackage{enumitem}
\usepackage{hyperref}
\usepackage{url}
\usepackage{booktabs}
\usepackage{multirow}
\usepackage{multicol}
\usepackage{adjustbox}
\usepackage{array}
\usepackage{caption}
\usepackage{amsmath}
\usepackage{algorithm}
\usepackage{algorithmic}
\usepackage{xcolor}
\usepackage{amssymb}

\usepackage{siunitx}
\usepackage{amsthm}
\usepackage[hang,flushmargin]{footmisc}
\usepackage{cryptocode}
\usepackage{threeparttable}

\usepackage{changes}
\usepackage{comment}

\usepackage[numbers,sort&compress]{natbib}

\def\BibTeX{{\rm B\kern-.05em{\sc i\kern-.025em b}\kern-.08em
    T\kern-.1667em\lower.7ex\hbox{E}\kern-.125emX}}

\usepackage{pifont}
\newcommand{\cmark}{\ding{51}}
\newcommand{\xmark}{\ding{55}}

\newcommand{\facc}[2]{{#1}{\scriptsize$\pm${#2}}}

\definecolor{darkred}{rgb}{0.55, 0.0, 0.0}

\newcommand{\PreserveBackslash}[1]{\let\temp=\\#1\let\\=\temp}
\newcolumntype{C}[1]{>{\PreserveBackslash\centering}p{#1}}
\newcolumntype{R}[1]{>{\PreserveBackslash\raggedleft}p{#1}}
\newcolumntype{L}[1]{>{\PreserveBackslash\raggedright}p{#1}}

\newcommand{\topic}[1]{\smallskip\noindent\textbf{#1}}

\newcommand{\SH}[1]{%
    \vspace{0.2em}{\noindent \color{red}\textbf{Sanghyun:} #1}\vspace{0.2em}}

\begin{document}

\date{}


\title{Publishing Efficient On-device Models Increases Adversarial Vulnerability}


\author{
    {\rm Sanghyun Hong, $^\dagger$Nicholas Carlini, $^\dagger$Alexey Kurakin } \vspace{0.2em} \\
    Oregon State University, $^\dagger$Google Brain \\
    { \small \tt sanghyun.hong@oregonstate.edu, \{ncarlini, kurakin\}@google.com }
} 

\maketitle


%
\begin{abstract}
Recent increases in the computational demands of deep neural networks (DNNs) 
have sparked interest in efficient deep learning mechanisms, e.g., quantization or pruning. 
These mechanisms enable the construction of a small, efficient version of commercial-scale models 
with comparable accuracy, accelerating their deployment to resource-constrained devices.

In this paper, we study the security considerations of publishing on-device variants of large-scale models.
We first show that an adversary can exploit on-device models
to make attacking the large models easier.
In evaluations across 19 DNNs,
by exploiting the published on-device models as a transfer prior, 
the adversarial vulnerability of the original commercial-scale models increases by up to 100x.
We then show that the vulnerability increases as the similarity between a full-scale and its efficient model increase.
Based on the insights, we propose a defense, \emph{similarity-unpairing}, 
that fine-tunes on-device models with the objective of reducing the similarity. 
We evaluated our defense on all the 19 DNNs 
and found that it reduces the transferability up to 90\% 
and the number of queries required by a factor of 10--100x. 
Our results suggest that further research is needed 
on the security (or even privacy) threats caused by publishing those efficient siblings.
\end{abstract}


\section{Introduction}
\label{sec:intro}

%
%
Deep neural networks (DNNs) are vulnerable to \emph{adversarial examples}~\cite{Intriguing}: 
%
an adversary can craft human-imperceptible perturbations to inputs
so that models make controlled mistakes.
Yet, conducting real-world adversarial attacks is challenging.
Many existing machine learning (ML) systems,
e.g., spam/abuse detection~\cite{SpamEvasion, NSFWCVLab, GoogleSafeSearch},
that carry incentives for potential adversaries,
are deployed with server-side models 
that have limited query access.
To fool such systems,
an attacker has to conduct black-box adversarial attacks~\cite{AutoZoom, P-RGF, BanditsTD, HybridBatch, DRMI}, 
which typically requires thousands of queries to craft a single adversarial example.
Additional safety mechanisms, 
\textit{e.g.}, banning malicious users,
also hinder the adversary's capabilities~\cite{StatefulDetection, BlackLight}

%
A recent trend in ML deployment is to publish DNN models to user devices.
Diverse efficient deep learning mechanisms, 
ranging from compression, 
e.g., quantization or pruning~\cite{FirstQuantization, PACT, BinaryConnect, XORNet, Pruning, LTH},
to designing new efficient architectures~\cite{MobileNet, SqueezeNet, NASNet, OFA},
have been proposed to bring commercial-scale server-side models
to devices with constrained resources.
%
%
Pushing on-device models
enables service continuity by making predictions offline 
and reduces the operational cost of service providers.
However, this \emph{also} makes it easier for an adversary 
to reverse-engineer these models and use them to their advantage.
Since the same providers likely construct on-device models
who trained the server-side counterparts,
they likely share many similarities, e.g.,
they are trained on the same (or very similar) data
or use similar architectures (but differ in size).

%

In this paper, we study the additional increase in vulnerability
caused by the common practices of efficient deep learning.
Specifically, we ask the questions:
%
\begin{quote}
    \emph{Does the publishing of on-device models make ML systems vulnerable to adversarial attacks?
          And if the answer is yes, then what can we do to remedy it?}
\end{quote}
%
Our work focuses on the black-box adversarial attacks as they exploit the technique, 
i.e., estimating the gradients of the target model from query outputs,
applicable to other security or privacy threats,
such as model extraction/inversion~\cite{ModelExtraction, ModelInversion}.

%
We begin by a comprehensive analysis
that characterizes the success of black-box adversarial attacks 
both with and without access to on-device models.
We examine 19 different pairs of models 
constructed from six different efficient deep learning mechanisms 
on two image classification benchmarks.
We find that adversarial examples crafted on on-device models transfer with up to 100\% success.
Moreover, an adversary 
requires $10$--$100\times$ fewer queries to fool the server model 
and have a higher attack success
if they have access to an on-device model.

%
We next study and characterize this increased vulnerability to understand why it occurs.
We find that (i) training on-device models longer reduces the vulnerability, 
and (ii) the more the difference between server and on-device models' architectures, 
the less vulnerability is.
This observation suggests a \emph{trade-off} 
between the security and the efforts in constructing on-device models
in efficient deep learning techniques.
%
We show that more accurate techniques induce a higher vulnerability.
We then propose three metrics to measure model similarity, 
and show that our metrics effectively predict
how vulnerable a model will be by having the on-device variant.

%
Informed by this characterization,
we propose a defense that allows training on-device models
while minimizing---but not completely eliminating---the increased vulnerability.
Our \emph{similarity-unpairing} defense fine-tunes the client model 
with the objective of making it dis-similar to the server-side model.
%
We comprehensively evaluate the effectiveness of 
similarity-unpairing in the network pairs we tested in our vulnerability analysis.
The adversarial examples crafted on
the fine-tuned models transfer 50--80\% less than the attacks with the original on-device model,
and we force the adversary to use 2--10$\times$ more queries to achieve similar or fewer successes in black-box attacks.
While promising, future work will be necessary to eliminate this increased vulnerability completely.

%
\topic{Contributions.}
We summarize our contributions as follows:
\begin{itemize}[nolistsep, topsep=0.2em]
    \item We present a new threat model 
    where an attacker exploits on-device variants of server-side models 
    to increase the adversarial vulnerability. 
    We formalize this as a security game and identify the goals for the attacker and defender.
    \item We show that the attacker can increase the vulnerability of server-side models significantly 
    by exploiting their on-device variants.
    The adversarial examples crafted on on-device models transfer at most 100\% to those models 
    while reducing the query complexity by 10--100$\times$.
    \item We further characterize the power of using on-device models 
    obtained by 
    efficient deep learning mechanisms.
    We find a trade-off between security and the efforts in constructing these on-device priors.
    \item We study three metrics that allow us to quantify how much two models are \emph{similar} 
    and evaluate their effectiveness in capturing the vulnerability.
    \item We propose a defense, similarity-unpairing, 
    that forms an objective function for reducing the similarities between on-device and server models. 
    We present three strategies to incorporate this defense into efficient deep learning and demonstrate their effectiveness.
\end{itemize}

\section{Background}
\label{sec:prelim}

We start by reviewing the paradigm of efficient deep learning and adversarial attacks and relates the two research fields.

%
\subsection{Efficient Deep Learning}
\label{subsec:efficient-deep-learning}

Efficient deep learning develops a wide range of mechanisms 
for increasing the efficiency of the final trained model 
while preserving its original accuracy.
This is a vast research field; we study the most popular techniques 
including training new efficient architectures~\cite{SqueezeNet, MobileNet, MobileNetV2, DenseNet} 
(possibly with neural architecture search~\cite{ProxylessNAS, AutoMLforEDL, NAS, NAS2}),
knowledge distillation~\cite{KD, QuestionableKD, IneffectiveKD} to transfer knowledge from a larger model into a smaller model,
quantization~\cite{FirstQuantization, PACT, BinaryConnect, LQNets, XORNet} to reduce the size of the model parameters on disk,
and pruning~\cite{Pruning, LTH} to reduce the total number of connections in a neural network.
We also consider 
a recent direction, \emph{once-for-all}, 
where a single model can be trained to support multiple different efficient scales at the same time~\cite{OFA}.
In \S\ref{subsec:exp-setup}, we provide details of these methods.

%
\subsection{(Black-box) Adversarial Attacks}
\label{subsec:adversarial-examples}


Adversarial examples \cite{Intriguing, Evasion} allow an adversary 
to arbitrarily change a neural network's predictions 
by adding small perturbations to its test-time inputs.
Many adversarial-example crafting algorithms~\cite{Evasion, FGSM, PGD, PapernotLimits, DeepFool, CNW, PGD} 
have been proposed to benchmark the worst-case sensitivity of 
a neural network to input perturbations in \emph{white-box} settings, 
where an adversary has full knowledge of the target neural network.
For a given test-time sample $(x, y)$, 
these attacks search for an input $x'$ 
that maximizes the loss between a model's output and the true label $\mathcal{L}(f_{\theta}(x), y)$ 
while its perturbations are bounded $||x' - x||_{\ell_p}$ to be human-imperceptible.

Our paper focuses on the more realistic \emph{black-box} settings, 
where an adversary can only access the target model $f_{\theta}$ 
through API accesses~\cite{Transferability, Delving, Zoo, DecisionBased, BanditsTD, P-RGF, HybridBatch, DRMI}, 
and cannot directly compute gradients on the model.
Black-box attacks introduce a new constraint: 
adversaries must minimize the total number of queries made to $f_{\theta}$.
Prior work presented two different strategies: 
\emph{transfer-based} and \emph{optimization-based} attacks.

\topic{Transfer-based attacks.}
Transfer-based attacks exploit the \emph{transferability} phenomenon of adversarial examples~\cite{Transferability}:
it turns out that if an adversary generates an adversarial example on one model (a ``surrogate model''),
this exact adversarial example often fool another remote target model.
%
%
%
Prior work~\cite{Delving} showed that the effectiveness of transfer-based attacks depends on how \emph{similar} a surrogate is to the target. 
%
%

%
%
%

\topic{Optimization-based attacks.}
If query access to the remote system is possible, early work
proposed techniques to first 
reconstruct a
similar model and then using that as a transfer source
\cite{LiQueryEfficient, PracticalBlackbox, DRMI}.
However, future 
optimization-based black-box attacks are often strictly more powerful.
%
They generate adversarial examples by defining an objective function and iteratively perturbing the input to solve that objective.
%

There are two general attack approaches.
Gradient estimation attacks~\cite{Zoo, Bhagoji, AutoZoom, NESAttack, P-RGF} approximate 
the target model's gradient from query outputs (\textit{i.e.}, softmax confidence scores),
and run standard white-box attacks using this estimated gradient.
%
%
Gradient-free attacks~\cite{GenAttack, Narodytska, SimBA, NAttack, Moon:ICML19, HopSkipJump} 
only rely on the model's hard decision (\textit{i.e.}, the predicted class), 
but often require (many) more queries to the target.
%
%
%
In this work, we focus on gradient estimation attacks as we aim to give the adversary every possible advantage when designing our defense.
%
%

\topic{Query attacks with transfer priors.}
A final direction unifies the two above attack techniques~\cite{BanditsTD, P-RGF}, 
and exploits transfer priors to perform a query attack.
In our experiments we use P-RGF~\cite{P-RGF} which works by
exploiting surrogate models available to the adversary.
We use as surrogates an on-device model that has been 
reverse-engineered by the adversary.

\section{Problem Formulation and Defense Goals}
\label{sec:problem}


%
\subsection{Threat Model}
\label{subsec:threat-model}

Black-box adversarial attacks are a far more practical threat today~\cite{dong2019cvpr, StealthyPorn, yu2020cloudleak, Tu2020CVPR, Hussain2021WACV}, so we focus on them in our paper.
For example, many \emph{abuse detection} tasks,
\textit{e.g.}, spam/phishing detection or not-safe-for-work (NSFW) detection, 
only allow query access to the classifiers running on servers.
For simplicity, 
we consider the task of \emph{image classification} 
because it is the domain that has seen most research on adversarial attacks.
However, our scenario could also be easily extended to other applications, \textit{e.g.}, text classification.

%
%
%

As mentioned 
in \S\ref{subsec:adversarial-examples}, 
black-box attacks can be made much more efficient 
when transfer priors are available.
We consider the specific scenario where \emph{defender} 
supplies the surrogate to the adversary ``for free'' 
by releasing an efficient on-device model 
that the adversary can use as the transfer prior.
%
%
%
%
We discuss the practical attack scenarios in Appendix~\ref{appendix:practical-attack-scenarios}.

%
%
\topic{Notation.}
We use the following notation 
throughout the paper.
\begin{itemize}[nolistsep, topsep=0.1em]
    \item \textbf{Adversary (A):} An entity who fools a model by constructing adversarial examples.
    \item \textbf{Model provider (P):} An entity that trains models, and then use them to provide services to general public.
    \item \textbf{Server-side model (${f_{\theta_s}}$):} The trained model available remotely on the server.
    \item \textbf{On-device model ($f_{\theta_o}$):} The efficient version of $f_{\theta_s}$ that is released publicly and is on user devices.
\end{itemize}
The provider \textbf{P} offers two types of models:
\begin{itemize}[nolistsep, topsep=0.1em]
    \item \textbf{White-box:} The adversary has complete access to the 
    parameters of the model, and can run arbitrary attacks; the on-device models
    fall into this category.
    %
    %
    \item \textbf{Black-box with limited API access:}
    The adversary is only allowed to access a query interface
    and cannot obtain gradients from the model,
    as happens in the server-side setting.
    For example, \textbf{P} is an image hosting provider 
    and uses a server-side model to detect NSFW images.
    The model's decision affects 
    whether the user will be further allowed to upload images or not. 
    \textbf{A} can only observe the outcome of uploading a few image. 
    %
\end{itemize}

\topic{Adversary objective and cost.}
The attacker receives an input $x$, has query access to the server-side model $f_{\theta_s}$, 
and generates an adversarial example $x'$ so that $f_{\theta_s}$ misclassifies $x'$.
%
In our threat model, 
\textbf{A} has complete white-box access to the on-device model $f_{\theta_o}$ and
is allowed to make a limited number of queries to $f_{\theta_s}$.
The attack \emph{cost} is measured as the number of times \textbf{A} queries to $f_{\theta_s}$ to craft $x'$.
In transfer-based attacks, \textbf{A} only uses $f_{\theta_o}$ for crafting; thus, the cost is zero.

%
\subsection{Attacker and Defender Security Game}
\label{subsec:security-game}

We begin by recalling the standard security game (denoted Game A), 
used throughout adversarial machine learning 
that defines the \emph{robust accuracy} of a machine learning model $f_{\theta_s}$.

\begin{enumerate}[label=(\arabic*), nolistsep, topsep=0.1em]
    \item The defender trains a 
    model $f_{\theta_s}$ with training algorithm $\mathcal{T}_1$.
    This model can be arbitrarily large and 
    is intended to reach state-of-the-art accuracy
    at the desired task.
    \item The defender grants the adversary query access to $f_{\theta_s}$.
    \item The adversary samples a benign example $x \gets \mathcal{X}$ from the data distribution,
    and generates an adversarial example $x'$ on $f_{\theta_s}$ by making queries to this model.
    The adversary \emph{fails} if $f_{\theta_s}$ correctly classifies $x'$;
    otherwise, adversary \emph{succeeds} and 
    we measure the cost of the attack as the number of times the adversary queries $f_{\theta_s}$.
\end{enumerate}

This security game changes 
when we focus on the problem of this paper and 
grant the adversary access to an on-device model 
that the defender has also constructed.
Under this setting, 
the new security Game B proceeds as follows:

\begin{enumerate}[label=(\arabic*), nolistsep, topsep=0.1em]
    \item The defender creates {\color{darkred}two models}: 
    $f_{\theta_s}$, using the same $\mathcal{T}_1$,
    {\color{darkred}and a new model $f_{\theta_o}$ with an efficient deep learning algorithm $\mathcal{T}_2$.}
    As before, $f_{\theta_s}$ is large and accurate; and this time,
    the on-device model $f_{\theta_o}$ should be small and efficient 
    for inference, but still achieve high accuracy.
    \item The defender grants the adversary query access to $f_{\theta_s}$
    and {\color{darkred}sends the adversary the on-device model $f_{\theta_o}$}.
    \item By exploiting {\color{darkred}the white-box access to $f_{\theta_o}$}, the adversary generates
     a new adversarial example $x''$.
     The \textbf{success} criteria and the \textbf{cost} metric remain unchanged;
     importantly, only $f_{\theta_s}$ needs to be fooled and 
     only queries to $f_{\theta_s}$ count against the cost.
\end{enumerate}

\newtheorem{definition}{Definition}
\begin{definition}
\normalfont The \emph{vulnerability increase} is defined by the difference in queries necessary to fool the model,
formally:
\[Pr[\mathcal{A}\,\, \text{succeeds at Game B}] - Pr[\mathcal{A}\,\, \text{succeeds at Game A}]. \]
\end{definition}
A more refined analysis can be made by considering the success rate and the reduction in queries necessary to succeed on average.
While we generally find these two to be highly correlated, 
in principle, they need not be, and 
so we make our definition with respect to success rate and 
use query count as an additional metric if the success rates are comparable.

\topic{Defender objective.}
The defender's objective is to develop 
a new training algorithm $\mathcal{T}_2'$
that minimizes the vulnerability increase.
Note here that $\mathcal{T}_1$ is a fixed function and can not be changed:
the server training algorithm is likely a complex setup 
that is designed to yield the highest quality 
$f_{\theta_s}$ possible, and 
it is unacceptable to reduce the quality of the model $f_{\theta_s}$ 
in order to permit releasing on-device models $f_{\theta_o}$.

Note that this game is only interesting 
if the server model $f_{\theta_s}$ is not trivially fooled from the beginning.
If the adversary, even without access to $f_{\text{o}}$, 
can win at the game with probability $\sim\!1.0$, 
then the total vulnerability can not increase significantly 
by having the transfer prior $f_{\text{o}}$.
However if this happens, 
then it is true that releasing the model will not cause harm,
because the server model is already trivial to evade.

%
\subsection{What Makes This Game Tractable?}

Over the past decade, 
there has been limited progress towards ``solving''
the problem of adversarial examples~\cite{FeatureSqueezing:NDSS18, PGD, TRADES:ICML2019, PixelDP:SNP2019, RandomizedSmoothing:ICML19, DetectingHard:ICMLWorkshop2021}.
The best approaches available 
either significantly harm accuracy~\cite{TRADES:ICML2019}, 
cause $100-1000\times$ increase in computations~\cite{PixelDP:SNP2019},
or do not scale to state-of-the-art models~\cite{PixelDP:SNP2019, RandomizedSmoothing:ICML19}.

We expect our problem formulation is one that can be reasonably solve because
\emph{the defender does not need to train models outright robust}; 
not to transfer attacks, not to query attacks, and certainly not to gradient-based attacks.
Instead, all the defender needs to do is \emph{not to make the situation worse}.

\topic{A trivial and uninteresting, but perfect solution.}
There is a straightforward solution to this problem:
the defender could train the on-device model 
using a completely different training setup, 
on a completely different training dataset,
using a completely different model architecture and hyper-parameters.
Because this is something the adversary could have done themselves,
the advantage of the above game is by definition $0$, and 
this would amount to a perfect solution to our problem.

There are two reasons
this solution is undesirable.
\begin{itemize}[nolistsep, topsep=0.1em]
    \item It considerably increases the defender's training complexity.
    The defender now must maintain two independent training setups,
    must fix bugs in two independent training algorithms, 
    must gather two different training datasets, 
    and in general must do everything twice.
    
    \item The on-device model will be less accurate 
    if trained from scratch than derived from the server model.
    One of the main benefits of deriving a small on-device model 
    from a larger pre-trained model is that 
    the resulting on-device model can be made more accurate.
    Training a smaller model completely from scratch 
    often does not reach the same accuracy levels.
    While in principle, the defender could entirely re-create the server model training setup and 
    then compress this model down to an on-device model 
    for the sole purpose of obtaining a high accuracy on-device variant, 
    this is an even larger complexity.
\end{itemize}

\topic{Goals.}
The objective of the defender, therefore, is to find a 
\emph{nontrivial} solution to this problem,
and minimize the increased vulnerability of the server model to
black-box adversarial examples without having to construct an entirely new and
complex training pipeline just to achieve this goal.

\section{Evaluating the Power of On-Device Priors}
\label{sec:exploit-compact-models}

We first study this problem from the adversary's perspective 
and quantify to what extent publishing on-device models 
increases the vulnerability of the server-side models to black-box adversarial attacks.
We outline our experimental setup and methodology (\S\ref{subsec:exp-setup}) 
followed by the vulnerability analysis (\S\ref{subsec:quantify-the-vulnerability}).
In \S\ref{subsec:characterize-the-vuln}, 
we characterize the vulnerability
considering the costs a victim requires to spend 
to obtain on-device models.
We further show there is no-free lunch in efficient deep learning (\S\ref{subsec:free-lunch}): 
lower-cost techniques to construct on-device models 
lead to the increase of the vulnerability 
more than higher-cost techniques.
We lastly discuss the security implications of our results and the desiderata for a defense.


%
\begin{table*}[t]
\centering
\begin{threeparttable}
\adjustbox{max width=\textwidth}{
    \begin{tabular}{c|lc|llcc|cc|rr}
    \toprule
    \multirow{2}{*}{\textbf{Dataset}} & \multicolumn{2}{c|}{\textbf{Original Model ($f_{\theta_s}$)}} & \multicolumn{4}{c|}{\textbf{On-device Model ($f_{\theta_o}$)}} & \multicolumn{2}{c|}{\textbf{Transfer-based ($\ell_{\inf}$)}} & \multicolumn{2}{c}{\textbf{Optimization-based}~\cite{P-RGF}} \\ \cmidrule{2-11}
     & \multicolumn{1}{c}{\textbf{Arch.}} & \textbf{Acc. (\%)} & \multicolumn{1}{c}{\textbf{Mechanism}} & \multicolumn{1}{c}{\textbf{Arch. (New)}} & \textbf{Need Training} & \textbf{Acc. (\%)} & \textbf{FGSM} & \textbf{PGD-10} & \textbf{\# Queries} & \textbf{FR (\%)} \\ \midrule \midrule
    \multirow{10}{*}{\rotatebox[origin=c]{90}{\textbf{CIFAR10}}} 
     & \multirow{10}{*}{ResNet50} & \multirow{10}{*}{\facc{91.0}{0.7}} & Baseline (as-is) & ResNet50 (\xmark) & \xmark & \facc{91.0}{0.7} & \facc{1.00}{0.00} & \facc{1.00}{0.00} & \facc{13.7}{\hspace{1.0em}1.7} & \facc{100}{0.0} \\
     &  &  & Baseline (a new) & ResNet50 (\xmark) & \cmark & \facc{91.0}{0.7} & \facc{0.17}{0.05} & \facc{0.66}{0.11} & \facc{863.6}{314.0} & \facc{82}{7.2} \\
     &  &  & Baseline (no $f_{\theta_o}$) & \multicolumn{1}{c}{-} & - & - & - & - & \facc{3047.2}{\hspace{0.5em}27.0} & \facc{28}{0.5} \\ \cmidrule{4-11} 
     &  &  & Quantization & ResNet50 (\xmark) & \xmark & \facc{91.0}{0.7} & \facc{1.00}{0.00} & \facc{1.00}{0.00} & \facc{13.9}{\hspace{1.0em}2.3} & \facc{100}{0.0} \\
     &  &  & Pruning & ResNet50 (\xmark) & \xmark & \facc{87.9}{0.9} & \facc{0.90}{0.01} & \facc{1.00}{0.00} & \facc{15.1}{\hspace{1.0em}4.1} & \facc{100}{0.1} \\
     &  &  & Distillation & ResNet18 (\xmark) & \cmark & \facc{87.9}{0.9} & \facc{0.18}{0.03} & \facc{0.70}{0.04} & \facc{752.3}{\hspace{0.5em}86.9} & \facc{85}{2.1} \\
     &  &  & Distillation & NASNet (\cmark) & \cmark & \facc{81.8}{1.3} & \facc{0.09}{0.02} & \facc{0.40}{0.02} & \facc{1713.4}{\hspace{0.5em}43.7} & \facc{62}{1.2} \\
     &  &  & NAS & NASNet (\cmark) & \cmark & \facc{84.7}{2.8} & \facc{0.11}{0.02} & \facc{0.46}{0.10} & \facc{1486.1}{367.1}& \facc{67}{8.7} \\
     &  &  & Manual-arch.& MobileNetV2 (\cmark) & \cmark & \facc{91.6}{0.9} & \facc{0.21}{0.03} & \facc{0.72}{0.03} & \facc{655.8}{118.5} & \facc{87}{2.7} \\
     &  &  & Manual-arch. & SqueezeNet (\cmark) & \cmark & \facc{89.1}{0.6} & \facc{0.11}{0.02} & \facc{0.45}{0.02} & \facc{1507.8}{121.5} & \facc{67}{3.5} \\ \midrule \midrule
    \multirow{18}{*}{\rotatebox[origin=c]{90}{\textbf{ImageNet}}} 
     & \multirow{8}{*}{ResNet50} & \multirow{8}{*}{\facc{75.3}{0.5}} & Baseline (as-is) & ResNet50 (\xmark) & \xmark & \facc{75.3}{0.5} & \facc{1.00}{0.00} & \facc{1.00}{0.00} & \facc{12.3}{\hspace{1.0em}0.1} & \facc{100}{0.0} \\
     &  &  & Baseline (a new) & ResNet50 (\xmark) & \cmark & \facc{75.3}{0.5} & \facc{0.34}{0.04} & \facc{0.74}{0.10} & \facc{649.9}{266.1} & \facc{87}{6.2} \\
     &  &  & Baseline (no $f_{\theta_o}$) & \multicolumn{1}{c}{-} & - & - & - & - & \facc{2558.9}{\hspace{0.5em}26.8} & \facc{38}{0.4} \\ \cmidrule{4-11} 
     &  &  & Quantization & ResNet50 (\xmark) & \xmark & \facc{75.3}{0.5} & \facc{1.00}{0.00} & \facc{1.00}{0.00} & \facc{12.2}{\hspace{1.0em}0.1} & \facc{100}{0.0} \\
     &  &  & Pruning & ResNet50 (\xmark) & \xmark & \facc{73.0}{0.0} & \facc{0.44}{0.29} & \facc{0.77}{0.14} & \facc{600.7}{364.3} & \facc{88}{7.8} \\
     &  &  & NAS & NASNet (\cmark) & \cmark & \facc{71.1}{1.9} & \facc{0.15}{0.02}& \facc{0.31}{0.05} & \facc{1952.4}{127.3} & \facc{54}{3.3} \\
     &  &  & Manual-arch. & MobileNetV2 (\cmark) & \cmark & \facc{70.8}{0.4} & \facc{0.16}{0.02} & \facc{0.35}{0.03} & \facc{1840.8}{\hspace{0.5em}66.5} & \facc{57}{1.9} \\
     &  &  & Manual-arch. & SqueezeNet (\cmark) & \cmark & \facc{50.2}{3.8} & \facc{0.08}{0.03} & \facc{0.28}{0.10} & \facc{1914.3}{429.6} & \facc{55}{8.0} \\ \cmidrule{2-11} 
     & \multirow{9}{*}{$^\dagger$OFA (4, 6, 7)} & \multirow{9}{*}{78.2} & Baseline (as-is) & OFA (4, 6, 7) (\xmark) & \xmark & 78.2 & 1.00 & 1.00 & 12.9 & 100 \\
     &  &  & Baseline (a new) & MobileNetV2 (\cmark) & \cmark & 72.0 & 0.27 & 0.53 & 1388.4 & 67 \\
     &  &  & Baseline (no $f_{\theta_o}$) & \multicolumn{1}{c}{-} & - & - & - & - & 2726.4 & 34 \\ \cmidrule{4-11} 
     &  &  & \multirow{6}{*}{Sub-networks} & OFA (4, 6, 3) (\xmark) & \xmark & 77.0 & 0.87 & 1.00 & 21.5 & 100 \\
     &  &  &  & OFA (4, 3, 3) (\xmark) & \xmark & 75.0 & 0.76 & 0.99 & 40.2 & 100 \\
     &  &  &  & OFA (2, 6, 7) (\xmark) & \xmark & 75.0 & 0.94 & 1.00 & 27.9 & 100 \\
     &  &  &  & OFA (2, 6, 3) (\xmark) & \xmark & 74.0 & 0.80 & 0.99 & 42.3 & 100 \\
     &  &  &  & OFA (2, 3, 7) (\xmark) & \xmark & 71.8 & 0.80 & 0.99 & 38.1 & 100 \\
     &  &  &  & OFA (2, 3, 3) (\xmark) & \xmark & 70.6 & 0.65 & 0.95 & 122.3 & 98 \\ \bottomrule
    \end{tabular}
}
\begin{tablenotes}
    \item $^\dagger$Note that we use the single pre-trained model that the original work~\cite{OFA} provides.
\end{tablenotes}
\end{threeparttable}
\caption{\textbf{Vulnerability to black-box attacks if transfer priors are on-device models,} evaluating on CIFAR10 and ImageNet.
We consider six different mechanisms for producing on-device models,
and we note the on-device architectures and whether or not the mechanism requires additional training.
Attacks are evaluated with FGSM, PGD-10, and P-RGF. 
We measure the relative fooling rate (RFR) for FGSM and PGD-10, and we report the \# Queries and the fooling rate (FR) for P-RGF.}
%
\label{tbl:effectiveness-of-priors}
\end{table*}

\subsection{Experimental Setup and Methodology}
\label{subsec:exp-setup}

\topic{Datasets.}
We run our experiments with two popular image classification benchmarks: CIFAR10~\cite{CIFAR10} and ImageNet~\cite{ImageNet}.
CIFAR10 is a 10-class classification dataset, with 50k training and 10k testing images of 32$\times$32 pixels.
ImageNet contains real-world images of 224$\times$224 pixels with 1000 categories.

\topic{Metrics.}
We use two metrics to quantify the vulnerability:
\begin{enumerate}[nolistsep, leftmargin=1.8em, label=(\arabic*)]
    \item 
    In transfer-based attacks, we design a new metric, relative fooling rate (RFR)%
    %
    \footnote{%
        RFR has the following advantages over the metrics proposed in prior work:
        (1) RFR is a standardized metric that enables comparing the effectiveness of different black-box attacks.
        It is not straightforward with the traditional metrics, e.g., an \emph{accuracy} of $f_{\theta_o}$ over $S^\prime$.
        (2) We encode the \emph{actual} attack success.
        Liang \textit{et al.}~\cite{Transfer:ICML'21} proposed a metric for quantifying the vulnerability, but we found that the metric won't capture the vulnerability in our scenarios.
        We include our further analysis in Appendix~\ref{appendix:other-metrics}.
    }, to measure the transferability.
    RFR is a standardized metric that allows us to compare the vulnerability across different datasets, models, and attacks:
    \begin{align*}
        \texttt{RFR}(f_{\theta_s}, S, S^{o}, S^{s}) = \frac{A(f_{\theta_s}, S) - A(f_{\theta_s}, S^{o})}{A(f_{\theta_s}, S) - A(f_{\theta_s}, S^{s})}
    \end{align*}
    where $A(f_{\theta}, S)$ denotes the accuracy of a model $f_{\theta}$ over a set of test-time samples $S$.
    $S^{s}$ and $S^{o}$ are the adversarial examples of $S$ crafted on $f_{\theta_s}$ and $f_{\theta_o}$, respectively.
    RFR computes how often the adversarial examples $S^{o}$ can be effective in causing an accuracy drop of $f_{\theta_s}$, 
    compared to the effectiveness of the white-box adversarial examples $S^{s}$ directly crafted on $f_{\theta_s}$.
    RFR will be close to 1, when most adversarial examples $S^{o}$ crafted on $f_{\theta_o}$ transfer to $f_{\theta_s}$; otherwise, RFR will be near 0.
    %
    %
    \item In optimization-based attacks, we measure the number of queries to the target 
    (\# Queries) required for crafting adversarial examples on average.
    We limit the number of queries an adversary can make to 4000 for each sample,
    following the prior work~\cite{HybridBatch, DRMI}. 
    If the adversary cannot fool the target after spending 4000 queries, 
    we stop crafting and count it as an attack failure.
    To quantify the attack success, 
    we also measure the fooling rate (FR). 
    FR is defined as the ratio of adversarial examples crafted by the attacker successfully fool the target.
\end{enumerate}

\topic{Methodology.}
We examine the vulnerability of server-side models to black-box attacks 
when an adversary can exploit on-device models as prior.
For transfer-based attacks, 
we craft adversarial examples using the on-device models and 
use them to attack the target (server-side) models.
We use two canonical adversarial-example crafting algorithms, FGSM and PGD-10.
In black-box optimization-based attacks, 
we use the P-RGF attack formulated by Cheng~\textit{et al.}~\cite{P-RGF}.
%
This attack requires a prior model, 
and for this we use each of the on-device models.
In each cases, 
we craft adversarial examples on the same 1000 samples randomly chosen from the test data.
We limit the perturbations to 8/255 pixels in $\ell_{\infty}$ norm 
(a standard value~\cite{PGD}) and 
fix the step-size to 2/255 pixels in the optimization process.
We run this experiment five times.

\topic{On-device Models.}
We examine six different mechanisms for constructing on-device models from the target $f_{\theta_s}$:
\begin{itemize}[nolistsep, topsep=0.1em, leftmargin=1.0em]
    \item \textbf{Quantization:} 
    We use 8-bit quantization to compress the target model. 
    Both the parameters and the activations are represented in 8-bit. 
    The on-device model has the size of one-fourth 
    and is computationally efficient 
    as it uses integer-only arithmetic, not floating-point operations.
    \item \textbf{Pruning:} 
    We use $\ell_{1}$-unstructured pruning, proposed by Li~\textit{et al.}~\cite{Pruning}. 
    We gradually increase the sparsity by 5\% from 0--100\% and stop 
    if the accuracy drop becomes more than 4\% of the target. 
    Pruned models we create have $\sim$50\% sparsity.
    \item \textbf{Knowledge Distillation (KD):} 
    We construct on-device models 
    with the knowledge distillation proposed by Hinton~\textit{et al.}~\cite{KD}. 
    We set the temperature to 20 and 
    the ratio between the main loss and the distillation loss to 1.0.
    \item \textbf{Neural Architecture Search (NAS):} 
    We further consider on-device models 
    that use the architectures constructed by NAS~\cite{NASNet} (e.g., NASNet). 
    This architecture has a smaller number of parameters and 
    achieves compatible accuracy. 
    We train them on the same training data from scratch.
    \item \textbf{Manually-designed Architectures:} 
    There are several architectures manually-designed 
    to reduce the number of parameters while achieving accuracy compatible with the target, 
    \textit{e.g.}, MobileNetV2 or SqueezeNet. 
    We train these models on the same training data from scratch.
    \item \textbf{Once-for-All (OFA) Models.} 
    We also consider the OFA model, proposed by Cai \textit{et al.}~\cite{OFA}, 
    where we can extract sub-networks---that have a smaller number of parameters---without training. 
    One can choose a sub-network from the pre-trained OFA based on the device's resource constraints.
\end{itemize}
We describe the experimental setup in detail in Appendix~\ref{appendix:experimental-details}.

%
\subsection{Vulnerability Increases When Using On-Device Models}
\label{subsec:quantify-the-vulnerability}

We show our results in Table~\ref{tbl:effectiveness-of-priors}.
The first three rows for each server model correspond to baselines:
\textit{Baseline (as-is)} directly uses the target as the transfer source,
and so is easiest to attack:
the transfer rate is $100\%$ because the models are identical,
and on average just 14 queries are required to fool the model.
Note that P-RGF uses the batch-size of 10 for querying.
\textit{Baseline (a new)} serves as our
perfect-but-uninteresting method described earlier: the adversary
trains a completely new model from scratch (using the same architecture);
this results in much worse transferability and $60\times$ higher query count than Baseline (as-is).
\textit{Baseline (no $f_{\theta_o}$)} does not use any on-device model 
for the black-box attacks, and thus performs the worst.
The remaining rows then exploit various on-device models constructed by using different efficient mechanisms.

\topic{Quantization and pruning increase vulnerability.}
To begin with, we study the effect of models constructed by quantization or pruning. 
In CIFAR10, the transfer-based attacks achieve 0.9--1.0 RFR in both FGSM and PGD, 
i.e., most adversarial examples crafted on on-device models successfully transfer to the target.
And for optimization-based attacks, the adversarial examples 
achieve 100\% attack success rate \emph{by making fewer than 15 queries on average}.
In contrast, without any prior, the attacker needs to query the target model 2559--3047 times on average, 
and even with $300\times$ as many queries, their resulting adversarial examples have a success rate of just 28--38\%.
This is consistent with prior work~\cite{WhyTransfer, P-RGF, HybridBatch, DRMI}, 
which has generally found that that optimization-based attacks require
thousands of queries to craft a single adversarial example.
However, our results show that just by having the quantized or pruned models as an on-device prior, 
the attacker reduces the query complexity (i.e., the cost) by two orders of magnitude.

\topic{Once-for-all paradigm also increases the vulnerability.}
We also observe that 
when the attacker has access to the sub-networks obtained from the OFA model, 
the vulnerability to black-box adversarial attacks increases significantly.
The last nine rows in Table~\ref{tbl:effectiveness-of-priors} presents the analysis for the OFA models.
The hyper-parameters for these models are denoted by a triple $(i,j,k)$ 
indicating the number of layers in each unit, the expansion ratio for each layer, and the kernel size (respectively).
We use the OFA (4, 6, 3) model as the target and 
choose six different sub-networks with smaller sizes as on-device models.
The vulnerability increased by these sub-networks 
is similar to (or even more) that caused by quantization and pruning.
In transfer-based attacks, we observe the RFR of 0.65--0.94 in FGSM and 0.95--1.00 in PGD-10, 
which is significantly higher than the baseline (a new) 
where we see 0.27 and 0.53 RFR in FGSM and PGD-10, respectively.
In optimization-based attacks, 
the attack requires 22$\times$--127$\times$ fewer queries than the baseline with no prior and 
11$\times$--64$\times$ fewer than the baseline that exploits an unrelated  MobileNetV2.
The attacker also achieves $\sim$100\% FR.

\topic{Models with architectures different from the target are less effective priors.}
The advantage of exploiting an on-device model decreases
if the model and the target have different architectures.
Using a different architecture requires training an on-device model from scratch;
thus, we compare our results with Baseline (a new).
In CIFAR10, NASNet or SqueezeNet on-device models achieve
lower RFRs (0.11 and 0.45 for FGSM and PGD-10, respectively) in transfer-based attacks 
and $2\times$ higher query complexity in optimization-based attacks than the baseline.
ResNet-like architectures, e.g., ResNet18 or MobileNetV2, 
when used as transfer priors, are more effective than NASNet and SqueezeNet.
Using MobileNetV2 on-device models achieves
almost the same effectiveness as the baseline.
In ImageNet, we observe similar results.

We also observe that knowledge distillation often increases 
vulnerability even when the two architectures are different.
In CIFAR10, a ResNet18 trained by knowledge distillation 
increases RFRs, requires fewer queries, and has a higher success rate.
The vulnerability eventually becomes similar to the baseline, 
where we use a ResNet50 on-device model trained from scratch.
However, when we use NASNet,
knowledge distillation does not increase the vulnerability further.
We hypothesize that this is because knowledge distillation forces 
the outputs of a teacher and a student to be \emph{similar} during training.
We will study this interaction further in \S\ref{subsec:similarity-between-the-two}.

%
\subsection{Exploitation in the Real World}
\label{subsec:real-world}

In \S\ref{subsec:quantify-the-vulnerability}, 
we find that an attacker can increase 
the risk of black-box 
attacks by exploiting on-device priors.
We now show that 
obtaining these priors is possible in practice.

Cloud providers offer many services 
where users can upload their datasets and train on-device models, 
e.g., Amazon SageMaker Edge or Google's AutoML Edge.
Those services are an attractive option for users with limited expertise in machine learning 
as they reduce the users' effort in optimizing their models for each device.
They automatically
train high-quality models within the resource limits.
In the meantime, service providers utilize 
efficient deep learning methods 
to minimize the cost of training on-device models.
If a user deploys them to edge, 
the attacker can reverse-engineer
and exploit them for crafting black-box adversarial examples.

\topic{Case study: SageMaker.}
We first attempted our attack on Amazon SageMaker Edge,
a service that provides both a larger server-side model along with an efficient edge model.
Here, we found that the on device models released by SageMaker Edge 
are \emph{functionally identical} to the large server-side models---they just use different
client-side libraries for deep learning computations 
optimized differently and use quantization.
This is equivalent to our \emph{quantization} setting.
And so in this case, we observe that 
both the transfer- and optimization-based attacks show 
$\sim$100\% success rate with less than 20 queries.
This result, while at the technical level is unsurprising because models are identical, 
demonstrates the potential pitfalls of releasing on-device models using off-the-shelf tooling.


\begin{table}[ht]
\centering
\adjustbox{max width=\linewidth}{
\begin{threeparttable}

    \begin{tabular}{@{}c|l|c|cc|cc@{}}
    \toprule
    \multirow{2}{*}{$f_s$} & \multicolumn{1}{c|}{\multirow{2}{*}{\textbf{On-device $f_c$}}} & \multirow{2}{*}{\textbf{Acc. ($f_c$)}} & \multicolumn{2}{c|}{\textbf{Transfer-}} & \multicolumn{2}{c}{\textbf{Optimization-}} \\ 
     &  &  & \textbf{FGSM} & \textbf{PGD10} & \textbf{\# Q} & \textbf{FR} \\ \midrule \midrule
    \multirow{9}{*}{\rotatebox[origin=c]{90}{\textbf{AutoML Vision}}}%
     & Baseline (no $f$) & 92\% & - & - & 3179.6 & 23\% \\ \cmidrule{2-7}
     & ResNet50 (P) & 91\% & 0.05 & 0.12 & 2832.3 & 33\% \\
     & ResNet50 (Q) & 88\% & 0.03 & 0.09 & 2879.3 & 30\% \\
     & ResNet18 & 89\% & 0.03 & 0.10 & 2775.4 & 34\% \\
     & NASNet & 89\% & 0.05 & 0.13 & 2844.1 & 33\% \\
     & MobileNetV2 & 93\% & 0.04 & 0.07 & 2957.8 & 29\% \\
     & SqueezeNet & 89\% & 0.04 & 0.12 & 2731.9 & 35\% \\ \cmidrule{2-7}
     & AutoML Edge$^\dagger$ & 93\% & 0.14 & 0.10 & 3254.4 & 20\% \\ \bottomrule
    \end{tabular}
    \begin{tablenotes}\footnotesize
    \item[*] We adapt our black-box attacks for this scenario.
    \end{tablenotes}
\end{threeparttable}
}
\caption{\textbf{Exploitation of the vulnerability in the real-world.} We first construct two models (server $f_s$ and on-device models $f_c$) using Google AutoML Vision, and we exploit the smaller, on-device model in attacking the cloud-side model.}
\label{tbl:cifar10-automl}
\end{table}

%
%
%
%
%
%

\topic{Case study: AutoML.}
To better understand how these attacks apply in practice, 
we next use Google's AutoML to construct a server-side model and an on-device model on CIFAR10.
In Google's AutoML, a user only has query access to the cloud model, 
i.e., the predicted class is only visible.
One can download the on-device model in TFLite~\cite{TFLite} format.
Unfortunately, TFLite does not allow to compute the gradients in its format, 
i.e., the attacker cannot directly apply FGSM or PGD on this model.
Instead, we assume that the attacker can reverse-engineer the model parameters from the format 
and use them to construct a surrogate $f_{\text{o'}}$, where we can approximate gradients.
We note that this adapted attack is much weaker than directly running the white-box attacks.
%
%
It enables us to assess the vulnerability in real-world scenarios.

Even with this weak, adapted adversary, Table~\ref{tbl:cifar10-automl} shows that 
an adversary can cause a similar or a better adversarial vulnerability than the baseline (no $f_{\theta_o}$).
In FGSM, the vulnerability can increase more than twice ($\sim$0.4$\rightarrow$0.14) 
if the attacker uses the approximated on-device model $f_{\text{o'}}$.
However, we also find that the vulnerability remains similar in PGD-10, 
and our adaptive optimization-based attack is ineffective.
In most PGD-10 cases, the accuracy drop caused by adversarial examples is low across the board.
The main reason is that the cloud models only allow using images in the integer format to query.
As we typically compute adversarial perturbations in a floating-point format, rounding them to integers can remove those small changes.
Under these constraints, the PGD-10 may already achieve the limit of the accuracy drop.


%
\begin{figure*}[ht]
    \minipage{0.25\textwidth}
      \includegraphics[width=\linewidth]{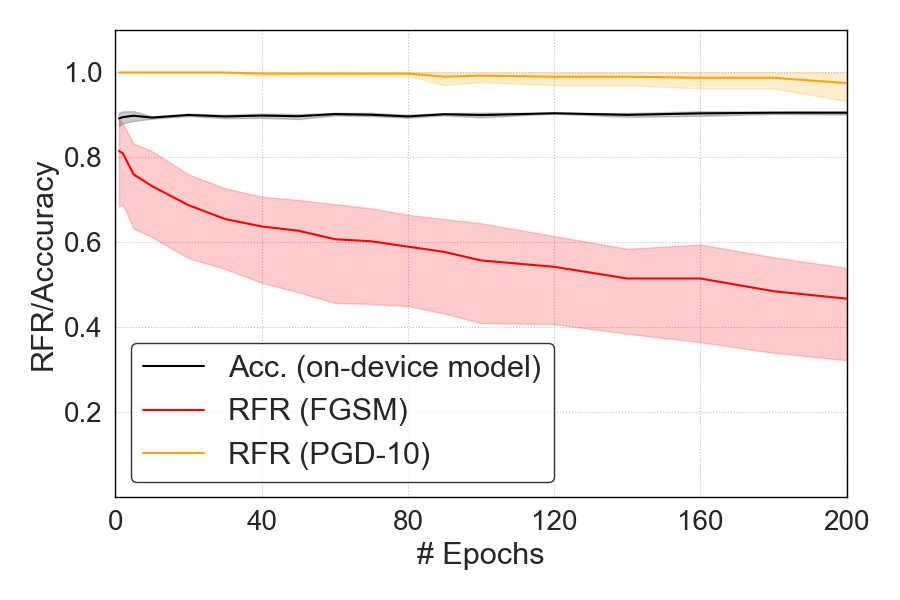}
    \endminipage
    \hfill
    \minipage{0.25\textwidth}
      \includegraphics[width=\linewidth]{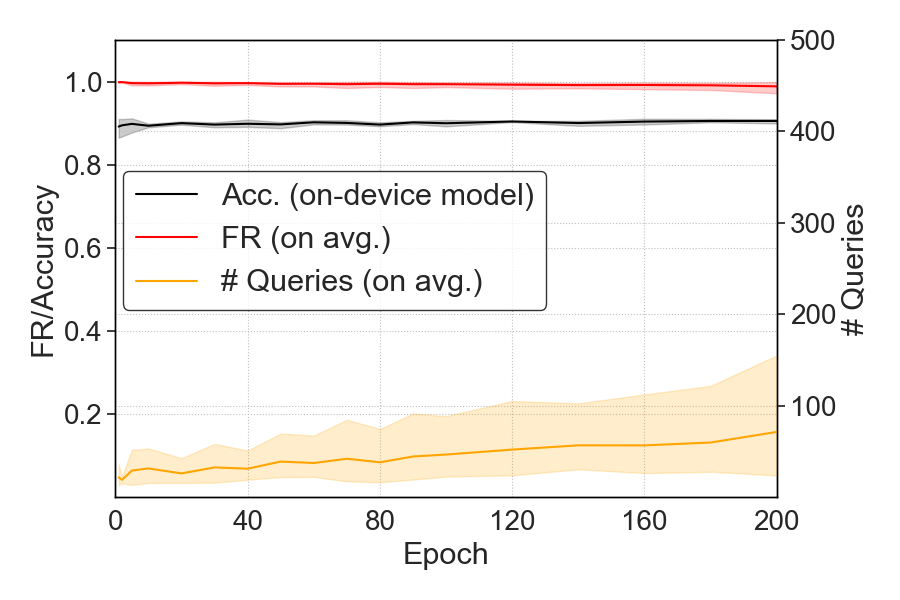}
    \endminipage
    %
    \hfill
    \minipage{0.245\textwidth}%
	  \includegraphics[width=\linewidth]{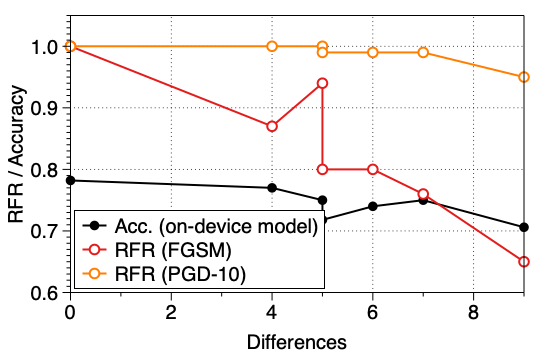}
    \endminipage
    \hfill
    \minipage{0.245\textwidth}%
	  \includegraphics[width=\linewidth]{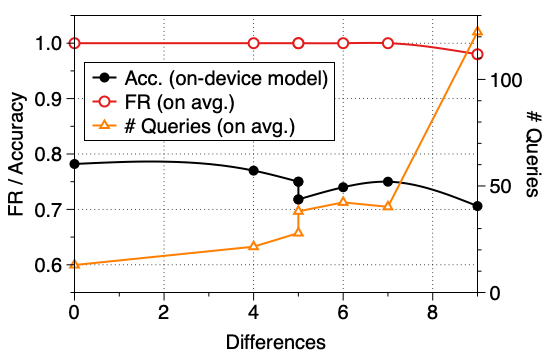}
    \endminipage
    \caption{\textbf{Vulnerability decreases as costs of constructing an on-device model increases.} We consider two different costs: the number of training iterations (\# Epochs) and the architectural difference between the target and the on-device model. In the left two figures, we fine-tune ResNet50 and measure the variability in the vulnerability. In the right two figures, based on the architectural difference we define between OFA networks, we measure how the vulnerability changes. As the costs increase, RFR decreases in transfer-based attacks, and the query efficiency and the success rate of the optimization-based attack decrease.}
    \label{fig:cost-security-analysis}
\end{figure*}

%
\subsection{Characterize the Vulnerability: Cost-Security Analysis}
\label{subsec:characterize-the-vuln}

Our analysis showed that 
efficient deep learning methods demand the training of an on-device model 
and utilize different architectures, resulting in less effective priors.
These two factors, i.e., the training iterations and the architectural differences, 
reflect how much effort we require to construct on-device models.
We conduct further analysis to characterize the interaction 
between the vulnerability to black-box adversarial attacks and the costs of building on-device models.

\topic{Longer training iterations reduces the vulnerability.}
To validate this hypothesis, 
we exploit fine-tuned models in transfer-based or optimization-based attacks.
We run fine-tuning of a ResNet50 model, 
pre-trained on CIFAR10, for 200 training iterations (epochs).
During training, we store the intermediate models in every 10 epochs.
We then use these models in crafting adversarial examples and 
measure the vulnerability of the original ResNet50 
models to black-box adversarial attacks.

The left two figures in Figure~\ref{fig:cost-security-analysis} show our results.
We first observe that the fine-tuning does not reduce the accuracy of the resulting models.
We also show that fine-tuning is effective in reducing the vulnerability of a weak transfer-based attack.
In the leftmost figure,
the RFR of FGSM decreases (0.82$\rightarrow$0.54) 
as the number of training epochs increases (0$\rightarrow$200).
However, fine-tuning is rendered ineffective 
when an adversary uses a stronger attack (PGD-10); it shows $\sim$1.0 RFR. 
The second figure from the left shows that 
the query complexity of the optimization-based attacks increases from 20 to 70 on average, 
meaning that more training iterations increase the attacker's cost, while the FR stays the same.

\topic{Larger architectural differences reduce the vulnerability.}
We evaluate this hypothesis using the OFA sub-network in Table~\ref{tbl:effectiveness-of-priors}.
It is challenging to compare the impact of the architectural difference because 
(i) models have different parameter values that can impact the vulnerability, and 
(ii) we have no metric to encode architectural differences.
We address the first by using those sub-networks---they share the same parameters derived from the OFA model.
Thus, we can minimize the impact of parameter differences.
We then define the architectural differences between the sub-networks 
by computing the $\ell_1$ distance between their configurations.
For example, the difference between the OFA (4, 6, 7) and OFA (2, 3, 7) models is $\lVert(4-2)+(6-3)+(7-7)\rVert_{\ell_1}\!=5$.
Note that this definition is not an ideal metric to measure the architectural differences---we just use it
as a proxy to characterize the interaction between the architectural difference and vulnerability.

We illustrate our results in Figure~\ref{fig:cost-security-analysis}.
Overall, we observe that the vulnerability decreases 
as the architectural difference increases, 
while the accuracy of the networks remains almost the same.
In transfer-based attacks, 
the RFR decreases from 1.00$\rightarrow$0.65 (FGSM) and from 1.00$\rightarrow$0.95 (PGD-10), respectively.
Like our previous analysis, 
choosing an architecture different from the target is more effective against a weak (FGSM) attack than the strong (PGD-10) attack.
%
In P-RGF, we find that the query complexity increases from 13$\rightarrow$122 on average, 
while the FR stays similar (100\%$\rightarrow$98\%).


\begin{figure*}[ht]
    \minipage{0.4\textwidth}
	  \includegraphics[width=\linewidth]{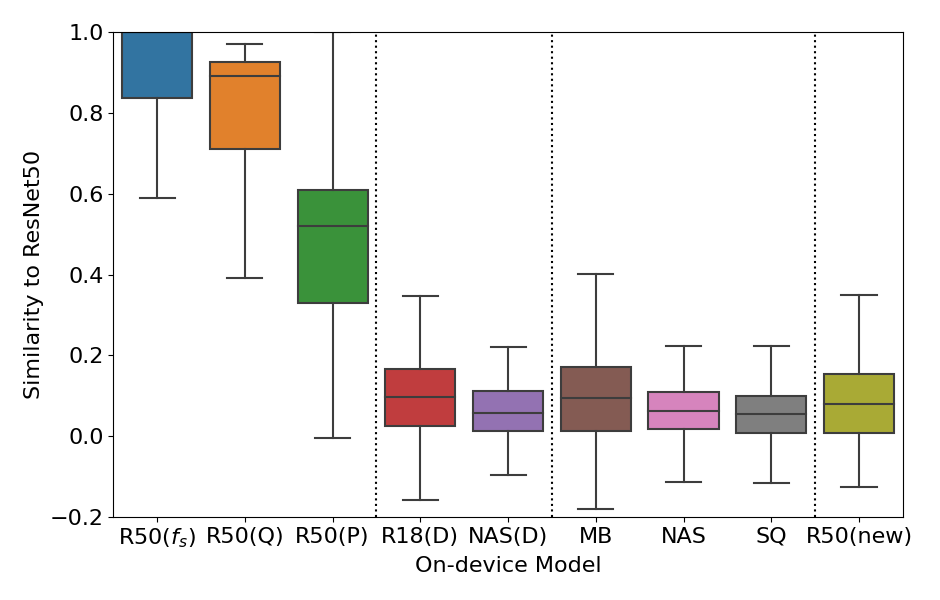}
    \endminipage
    \hfill
    \minipage{0.294\textwidth}
	  \includegraphics[width=\linewidth]{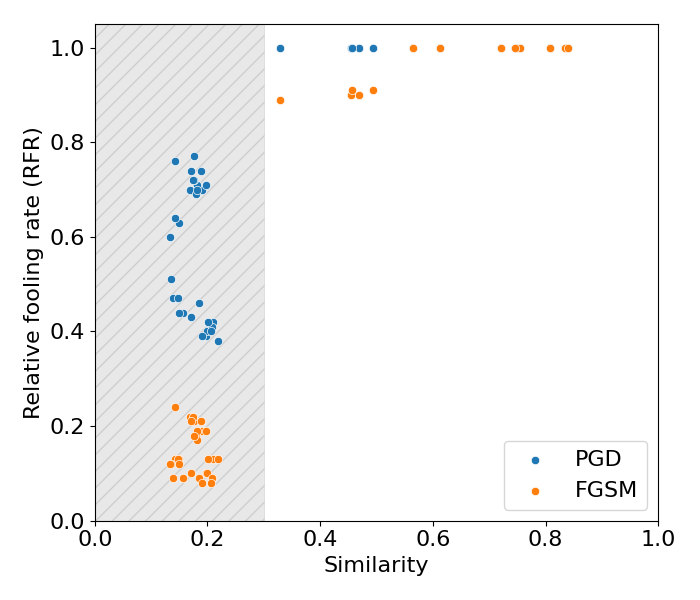}
    \endminipage
    \hfill
    \minipage{0.294\textwidth}%
	  \includegraphics[width=\linewidth]{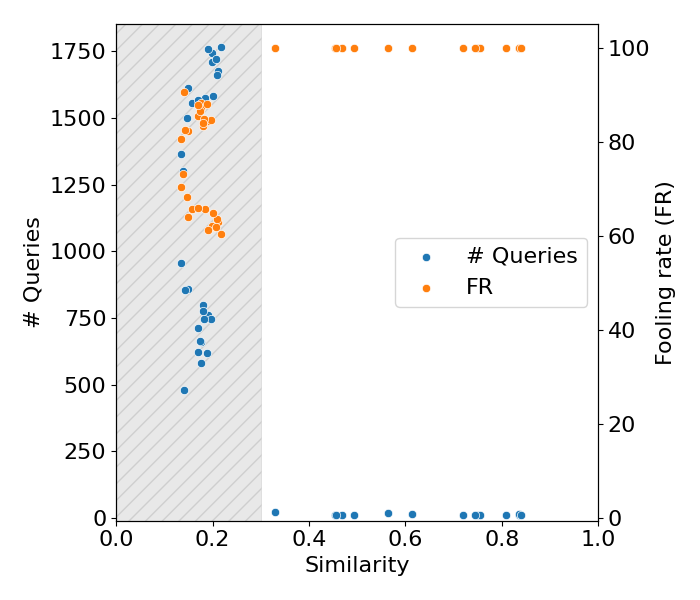}
    \endminipage
    \caption{
    \textbf{Our similarity metric accurately predicts the vulnerability to black-box adversarial examples.} We measure the similarities between the target (ResNet50) and the on-device models from Table~\ref{tbl:effectiveness-of-priors} at the input-gradients (left) on CIFAR-10. We plot the interaction between the vulnerability to black-box adversarial attacks and the similarity in the right two figures.}
    \label{fig:similarities}
\end{figure*}

%
\subsection{Security Implications}
\label{subsec:free-lunch}

\noindent
Our analysis shows a trade-off in efficient deep learning.
In particular, 
we look at the interaction between 
the security and the efforts in constructing on-device models.
If we put more effort into deriving on-device models 
(in terms of the training iterations and designing architectures), 
the vulnerability decreases.
Otherwise, if we use cheaper mechanisms like quantization and pruning, 
the resulting on-device models increase the vulnerablility of the server-side model.


\smallskip
\noindent \textbf{Desiderata.}
We summarize three conditions we would like out of an efficient deep learning training mechanism.
\begin{enumerate}[label=(\arabic*), noitemsep, topsep=0.em]
    \item \textit{No accuracy drop.} A mechanism will be ideal if the resulting on-device models have a compatible accuracy.
    \item \textit{Minimize the training cost.} Mechanisms that do not require training on-device models from scratch are better.
    \item \textit{Reduce the vulnerability.} 
    The vulnerability caused by on-device models should be less than the baselines in Table~\ref{tbl:effectiveness-of-priors}.
\end{enumerate}

%

%
\section{Similarity Unpairing Reduces Vulnerability}
\label{sec:vulnerability-reduction}

We now study the problem from the eyes of the defender, 
having shown in the prior section that 
without applying defensive techniques, 
there is a trade-off between the computational demands of efficient deep learning mechanisms and 
the security risk of a target model's to black-box adversarial attacks.
If the defender choose more computationally efficient techniques 
(e.g., pruning, quantizing, or the ``once-for-all" strategy) 
for constructing on-device models, 
it will increase the vulnerability to the black-box attacks.
However, mechanisms that train models from scratch do not increase vulnerability
(because the adversary could have done it them self)
but are computationally expensive and are also less accurate.

Here, we answer the question: 
can we reduce the vulnerability to black-box attacks further in an efficient way?
We propose a straightforward defense, \emph{similarity-unpairing}, 
that can actually achieve this by fine-tuning a model 
to reduce the similarities between $f_{\theta_s}$ and $f_{\theta_o}$.
To this end, we first explore metrics that quantify the similarity between the two models.
We measure how well the similarity metrics encode the vulnerability increase to black-box attacks.
We then develop a novel objective function 
that decreases the similarity between the two models $f_{\theta_s}$ and $f_{\theta_o}$ 
by fine-tuning the on-device model $f_{\theta_o}$.
We finally evaluate the effectiveness of our defense 
in reducing the vulnerability to black-box adversarial attacks 
when the attacker exploits the fine-tuned model $f_{o^\prime}$.

%
\subsection{Similarity Metrics to Quantify the Vulnerability}
\label{subsec:similarity-between-the-two}


We develop similarity metrics to quantify to what extent models produce similar classifications
in the hope that they will allow us to train models to be different as a defense.
We identify two locations in a model to measure similarity: activations and outputs (i.e., logits).
We additionally compute the gradients at the input space for the same test-time sample 
to capture the similarity of the two models' loss surfaces.
Using the three metrics (activations, outputs, and input-gradients), 
we measure the similarities between models in Table~\ref{tbl:effectiveness-of-priors}.

\topic{Metrics.}
We utilize the cosine similarity loss $C_s$ 
as a metric to quantify how much outputs obtained from each of the three locations are similar.
The loss value will be 1 when they are similar; otherwise, -1.
We compute the loss at each location as follows
(see the computation details in Appendix~\ref{appendix:sim-metrics}):
\begin{enumerate}[label=(\arabic*), nolistsep, topsep=0.1em]
    \item \textit{Output:} We compute the loss between the logits.
    \item \textit{Input-gradients:} We compute the loss between the input-gradients
    (i.e., the gradients computed on the same input with respect to two different models).
    \item \textit{Activations:} We compute the similarity between the activations of a model before the classification head (\textit{i.e.}, the latent representations), commonly regarded as features.
\end{enumerate}

\topic{Results.}
Figure~\ref{fig:similarities} shows our results in CIFAR10.
We compute the similarities between the target ResNet50 (R50) and 
the seven different models we construct in Table~\ref{tbl:effectiveness-of-priors}.
We use the same 1{,}000 samples used for crafting adversarial examples.

The leftmost figure shows the distribution of similarities observed at the input-gradients.
We find that the input-gradients reflect the vulnerability better than the two other metrics, 
i.e., logits and activations (see Appendix~\ref{appendix:limit-in-activation-sim}).
Quantization and pruning lead to on-device models with the highest similarities.
Perhaps unsurprisingly, we find that the similarity decreases 
as we use the mechanisms that require training a new model from scratch:
ResNet18 (D), NASNet (D), MobileNetV2, NASNet, SqueezeNet, and ResNet50 (a new),
confirming our observation in \S\ref{subsec:characterize-the-vuln}.
We further observe that, 
compared to the baseline where we train a new ResNet50 from scratch,
the architectural differences decrease the similarity further.


\begin{table*}[t]
\centering
\adjustbox{max width=\linewidth}{
    \begin{tabular}{@{}c|cccSc|cccSc|cccSc@{}}
        \toprule
        \textbf{Objective} & \multicolumn{5}{c|}{\textbf{Penalize Output-level Sim.}} & \multicolumn{5}{c|}{\textbf{Penalize Input-level Sim.}} & \multicolumn{5}{c}{\textbf{Penalize Feature-level Sim.}} \\ \midrule
        \multirow{2}{*}{\textbf{$\lambda$}} &  & \multicolumn{2}{c}{\textbf{Transfer-}} & \multicolumn{2}{c|}{\textbf{Optimization-}} &  & \multicolumn{2}{c}{\textbf{Transfer-}} & \multicolumn{2}{c|}{\textbf{Optimization-}} &  & \multicolumn{2}{c}{\textbf{Transfer-}} & \multicolumn{2}{c}{\textbf{Optimization-}} \\ \cmidrule(l){3-6} \cmidrule(l){8-11} \cmidrule(l){13-16}
         & \textbf{Acc.} & \textbf{FGSM} & \textbf{PGD10} & \textbf{\# Q} & \textbf{FR} & \textbf{Acc.} & \textbf{FGSM} & \textbf{PGD10} & \textbf{\# Q} & \textbf{FR} & \textbf{Acc.} & \textbf{FGSM} & \textbf{PGD10} & \textbf{\# Q} & \textbf{FR} \\ \midrule \midrule
        0.0 & 91\% & 0.84 & 1.00 & 46.2 & 99\% & 92\% & 0.83 & 1.00 & 29.1 & 100\% & 92\% & 0.83 & 1.00 & 30.1 & 100\% \\ \midrule
        0.001 & 91\% & 0.84 & 1.00 & 26.8 & 100\% & 91\% & 0.85 & 1.00 & 26.1 & 100\% & 92\% & 0.83 & 1.00 & 30.1 & 100\% \\
        0.01 & 91\% & 0.83 & 1.00 & 22.6 & 100\% & 91\% & 0.83 & 1.00 & 40.3 & 100\% & 92\% & 0.83 & 1.00 & 36.3 & 100\% \\
        0.1 & 91\% & \textbf{0.62} & \textbf{0.61} & 819.1 & 82\% & 91\% & 0.76 & 1.00 & 135.3 & 100\% & 91\% & 0.83 & 1.00 & 26.2 & 100\% \\
        1.0 & 90\% & 0.66 & 0.67 & 664.4 & 86\% & 91\% & \textbf{0.08} & \textbf{0.09} & \textbf{3411.9}\enspace\enspace\enspace & \textbf{15\%} & 91\% & 0.86 & 1.00 & 17.4 & 100\% \\
        10.0 & 90\% & 0.64 & 0.62 & \textbf{841.7}\enspace\enspace & \textbf{80\%} & 72\% & 0.22 & 0.86 & 191.9 & 97\% & 91\% & 0.85 & 1.00 & 24.1 & 100\% \\ \bottomrule
    \end{tabular}
}
\caption{\textbf{Effectiveness of our similarity unpairing defense.} We measure the vulnerability of $f_{\theta_s}$ to black-box adversarial examples when the attacker uses the model $f_{\theta_{c'}}$ fine-tuned with our similarity-unpairing loss. We use the ResNet50 model $f_{\theta_s}$ trained on CIFAR10. The cases where the attacker is the least successful are highlighted in bold.}
\label{tbl:effectiveness-of-unpairing}
\end{table*}




In the right two figures,
we plot the interaction between 
the vulnerability to black-box adversarial attacks 
and the similarity measured in the input-gradients.
We compute them with the models 
we examine previously in Table~\ref{tbl:effectiveness-of-priors}.
We first show that the input-gradient similarity directly predicts how easy it is to perform a black-box attack:
the model pair that is most similar is the best transfer source, 
and the least similar model pair is the worst transfer source.
And in query attacks (the right figure), 
as similarity increases, the query efficiency and the FR also increase.
We also find that decreasing the input-gradient similarity below 0.4 reduces the vulnerability significantly.
In transfer-based attacks, 
compared to the cases with a similarity score over 0.3,
the RFR of FGSM and PGD-10 decrease to 0.1--0.25 and 0.3--0.8 (respectively).
In optimization-based attacks,
the FR decreases to 55--90\%,
and the number of queries required increases to 1000--1600.
This connection gives the insight to develop techniques 
that reduce the vulnerability of black-box adversarial attacks 
in constructing models with efficient deep learning algorithms.

%
\subsection{Similarity Unpairing Objective}
\label{subsec:defense-setup}

We present our defense: \emph{similarity unpairing}.
The intuition for this defense is straightforward.
As shown above,
while we would like to directly convert the model $f_{\theta_s}$ 
into an efficient model $f_{\theta_o}$, 
this causes potential harm 
because an adversary can make use of the input-gradient similarity between $f_{\theta_o}$ and $f_{\theta_s}$.
So to prevent this, we fine-tune a model $f_{\theta}$ and
convert it into a new model $f_{{\theta'}}$ that
has similar accuracy, but a different input-gradient landscape.
Formally,
to reduce the similarity, 
we fine-tune the model on the following objective:
\begin{align*}
    \mathcal{L}_{ours} = \mathcal{L}_{\text{xe}} \big(f'(x), y \big) 
        + \sum_{i \in S} \lambda_{i} \cdot \mathop{\mathbb{E}_{\mathcal{D}}} \big[ {C_s}^{i}\big( x, y, f_{\theta}, f_{\theta'} \big) \big]
\end{align*}
where $\mathcal{L}_{\text{xe}}$ is the cross-entropy loss, 
${C_s}^i$ is the $i$-th similarity function we define in \S\ref{subsec:similarity-between-the-two}, 
$f_{\theta}$ and $f_{\theta'}$ are the original and fine-tuned models, 
and $\lambda_i$ is the hyper-parameter controlling the two loss terms.
We fine tune for 20 epochs 
(\textit{i.e.}, just $10\%$ of the total steps to train the server-side model), 
using the same set of hyper-parameters we use for training.
We consider only one similarity ${C_s}^i$ at a time, 
but we combine them together later on for reducing the vulnerability further.
We also train $f_{\theta'}$ from scratch, 
but this does not offer better protection.
We include those additional results in Appendix~\ref{appendix:additional-results-sim-reduction}.

\topic{Single objective at a time.}
Table~\ref{tbl:effectiveness-of-unpairing} shows the effectiveness of penalizing each similarity individually in reducing the vulnerability to black-box adversarial examples.
We fine-tune the ResNet50 model on CIFAR10 and then evaluate as before, using the fine-tuned model to craft adversarial examples on the target model, setting $\lambda_{i}$ in 0.001--10.0 to control the impact of our unpairing loss.
In each case, we measure the accuracy of the fine-tuned model $f_{\theta_{c'}}$ and the vulnerability metrics we define.
The baseline $\lambda_{i}=0$ is included as the first row.

We find penalizing the outputs and the input-gradients effectively reduces the vulnerability to black-box adversarial examples.
If we decrease the output-level similarity, the RFRs of the transfer-based attacks reduce from 0.84$\rightarrow$0.62 (FGSM) and 1.00$\rightarrow$0.61.
Additionally, the query complexity increases from 46.2 (baseline) to 841.7, and the success rate decreases from 100\% to 80\%.
Penalizing the similarity of the input-gradients is more effective than reducing the output-level similarity.
Here, in the transfer-based attacks, the RFRs are decreased to below 0.09 for both FGSM and PGD-10, with the query attack complexity increasing from 29.1 to 3411.9, and the FR reduces from 100\% to 15\%.
In both cases, we further preserve the accuracy of the fine-tuned models compared to the baselines.
However, we find that penalizing the activation-level similarity is not effective in reducing the vulnerability.


% ...
\begin{table}[ht]
\centering
\adjustbox{max width=\linewidth}{
    \begin{tabular}{@{}SS|c|cc|rc@{}}
    \toprule
    \multicolumn{2}{c|}{\textbf{Objectives}} & \multicolumn{5}{c}{\textbf{Penalize Multiple Sim.}} \\ \midrule
    \textbf{Output.} & \textbf{Input.} & & \multicolumn{2}{c|}{\textbf{Transfer-based}} & \multicolumn{2}{c}{\textbf{Optimization-based}} \\ \cmidrule(l){3-7} 
    \multicolumn{1}{c}{\textbf{$\lambda_1$}} & 
    \multicolumn{1}{c|}{\textbf{$\lambda_2$}} & \textbf{Acc.} & \textbf{FGSM} & \textbf{PGD10} & \multicolumn{1}{c}{\textbf{\# Q}} & \textbf{FR} \\ \midrule
    0.01 & 0.01 & 91\% & 0.25 & 0.87 & 291.1 & 95\% \\
    0.01 & 0.1 & 91\% & 0.23 & 0.83 & 344.2 & 96\% \\
    0.01 & 1.0 & 91\% & 0.05 & 0.08 & 2922.7 & 32\% \\ \midrule
    0.1 & 0.01 & 92\% & 0.24 & 0.79 & 381.2 & 94\% \\
    0.1 & 0.1 & 92\% & 0.20 & 0.77 & 491.5 & 91\% \\ 
    0.1 & 1.0 & 91\% & 0.06 & 0.12 & 2647.9 & 39\% \\ \midrule
    1.0 & 0.01 & 91\% & 0.18 & 0.64 & 564.4 & 91\% \\
    1.0 & 0.1 & 91\% & 0.18 & 0.65 & 567.5 & 91\% \\
    \hspace{1.0em}\textbf{1.0} & \hspace{0.6em}\textbf{1.0} & \textbf{93\%} & \textbf{0.06} & \textbf{0.08} & \textbf{2835.5} & \textbf{34}\% \\ \bottomrule
    \end{tabular}
}
\caption{\textbf{Combining multiple unpairing objectives.} We measure the vulnerability when we fine-tune on-device models while decreasing both the outputs and the input-gradients similarity. We highlight the least vulnerable case in bold.}
\label{tbl:combine-multiple-objectives}
\end{table}



\topic{Combining multiple unpairing objectives.}
We further examine if combining multiple objectives simultaneously can reduce the vulnerability more.
We fine-tune the ResNet50 model while reducing the similarities at the outputs and the input-gradients.
The hyper-parameters $\lambda_1$ and $\lambda_2$ are the weights for the output similarity loss and input-gradient similarity loss.
We vary them in 0.01--1.0.
(We exclude the activation-level similarity as it is ineffective.)

Table~\ref{tbl:combine-multiple-objectives} shows these results.
Combining multiple objectives reduces the vulnerability only slightly: if we set both $\lambda$s to 1.0, the RFRs of FGSM and PGD-10 become 0.06 and 0.08, just $\sim$0.02 less than penalizing only the similarity between input-gradients.
%
%
%
Therefore, for the remainder of our experiments, we consider just one loss penalty.

\section{Safe and Efficient Deep Learning\\with Similarity Unpairing}
\label{sec:defense-evaluations}

In \S\ref{subsec:defense-setup}, 
we show that our similarity unpairing 
would reduce the vulnerability of server-side models 
if the adversary had direct access to the fine-tuned model.
However, these fine-tuned models are still large (server-scale) models 
that can not be used on devices directly.
Here, we combine our defense with efficient deep learning mechanisms 
to construct efficient on-device models with reduced vulnerability.
We compare three strategies to
construct efficient and safe on-device models:
%

\begin{enumerate}[label=(\arabic*), nolistsep]
    \item \textit{Finetune first:} 
    We first fine-tune the model $f_{\theta_s}$ and apply efficient deep learning mechanisms, 
    \textit{e.g.}, the victim fine-tunes a pre-trained model and then applies quantization.
    \item \textit{Finetune last:} 
    We can also apply efficient deep learning mechanisms first to construct the on-device model $f_{\theta_o}$ 
    and then fine-tune it to reduce the vulnerability, 
    \textit{e.g.}, the victim constructs a quantized version of $f_{\theta_s}$ 
    and then fine-tunes the resulting model $f_{\theta_o}$ with our objective function.
    \item \textit{Jointly finetune:} 
    We can further develop novel efficient deep learning mechanisms 
    that return on-device models with reduced vulnerability 
    by incorporating our similarity-unpairing objective into their algorithms, 
    \textit{e.g.}, we add our objective to the distillation loss. 
\end{enumerate}
\vspace{0.2em}      


\begin{table}[ht]
\centering
\adjustbox{max width=\linewidth}{
    \begin{tabular}{@{}l|ccc@{}}
        \toprule
        \multicolumn{1}{c|}{\textbf{Mechanism}} & \textbf{Finetune first} & \textbf{Finetune last} & \textbf{Jointly finetune} \\ \midrule \midrule
        Quantization & \cmark & \cmark & \cmark \\
        Pruning & \cmark & \cmark & \cmark \\
        Distillation & \cmark & \cmark & \cmark \\
        New-arch. & \xmark & \cmark & \xmark \\
        OFA & \cmark & \cmark & \xmark \\ \bottomrule
    \end{tabular}
}
\caption{\textbf{Applicability of similarity-unpairing.} We show the applicability of our defense to six efficient deep learning mechanisms that our work examines. \cmark{} indicates that we can take the strategy, while \xmark{} means that it is not applicable.}
%
\label{tbl:defense-strategy}
\end{table}

Table~\ref{tbl:defense-strategy} shows the applicability of these three strategies 
to the efficient deep learning mechanisms that we examine.
The \emph{finetune-first} (FF) strategy is applicable when a mechanism uses a pre-trained model.
If the victim constructs on-device models by designing new architectures,
\textit{e.g.}, NAS, or manually-designed architectures, this strategy is not applicable.
The \emph{finetune-last} (FL) strategy is always applicable 
as one can fine-tune pre-trained on-device models with our defense objective.
We can apply the \emph{jointly-finetune} (JL) strategy 
when an efficient deep learning mechanism involves an optimization process.
The victim can add our similarity-unpairing loss 
as an additional objective for the process.
%

%
\subsection{Strawman Solution: Robust Server-side Models}
\label{subsec:robust-models}

Before we test whether our similarity unpairing reduces the vulnerability,
we examine a strawman solution that first constructs an adversarially-robust server-side model
and reduces its size so that on-device models are somewhat robust.

\topic{Methodology.}
We first run adversarial training (AT)~\cite{PGD}%
---a standard approach to train adversarially-robust models---on ResNet50.
We run AT with PGD-7
that has the perturbation bound of 8/255 pixels 
in $\ell_{\infty}$-norm and the step-size of 2/255 pixels.
We then utilize the six mechanisms we previously used 
to construct on-device models in \S\ref{subsec:characterize-the-vuln}.
Using the on-device models we construct as transfer priors,
we perform black-box adversarial attacks on the robust ResNet50.

\topic{Results.}
Due to the space limit, we summarize our results 
in Appendix~\ref{appendix:adversarial-training} and~\ref{appendix:robust-architecture}.
As we reviewed in \S\ref{sec:problem},
robust models suffer from the utility loss%
--their accuracy is 10--20\% less than the undefended models.
%
%
%
%
In transfer-based attacks, we observe that 
robust server-side models cannot reduce the vulnerability increase.
Quantization and pruning lead to the RFR of 0.8--1.0.
In the NASNet, MobileNetV2, and SqueezeNet cases,
we observe the RFR further increases to 0.5--0.6;
they were 0.1--0.2 in our results with non-robust server-side models.
Nevertheless, we find that robust server-side models 
reduce the vulnerability increase in query-based attacks.
Even in the most vulnerable cases, 
\textit{i.e.}, baseline (as-is), quantization, and pruning,
the FR is 53--64\% and, 
the number of queries needed for crafting a successful adversarial example is 1476--1886.
In remaining cases, the FR is significantly reduced to $\sim$25\%,
and the attacker requires $\sim$3000 queries to construct an adversarial example.
%
%
It is therefore necessary to construct defenses 
that allow the release of safe on-device models 
until the trade-off of training truly robust models is deemed acceptable.

%
\subsection{Similarity Unpairing with Quantization}
\label{subsec:defense-quantization}

\noindent \textbf{Methodology.}
%
%
To apply our defense to the FF setting,
we fine-tune a ResNet50 $f_{\theta_s}$ with our similarity-unpairing loss for 10--30 epochs and then quantize the model.
In the FL setting, we first quantize the pre-trained ResNet50 in 8-bit and then fine-tune for 10--30 epochs.
For finetuning (JL) we develop a quantization-aware training (QAT) scheme by minimizing:
\begin{align*}
	\mathcal{L}(f_{\theta_s}, x, y) = \mathcal{L}_{\text{xe}}(f_{\theta_s}(x), y) \
		+ \alpha \cdot \mathcal{L}_{ours}(f_{\theta_o}(x), y)
\end{align*}
where $f_{\theta_o}$ is the 8-bit representation of $f_{\theta_s}$, 
and $\alpha$ is the hyper-parameter between the two loss-terms.
We set $\alpha$ to 1.0.


%
\begin{table}[h]
\centering
\adjustbox{max width=\linewidth}{
    \begin{tabular}{@{}cccccSc@{}}
        \toprule
        \multirow{2}{*}{\textbf{Dataset}} & \multirow{2}{*}{\textbf{Strategy}} & \multirow{2}{*}{\textbf{\begin{tabular}[c]{@{}c@{}}Acc.\\ (8-bit)\end{tabular}}} & \multicolumn{2}{c}{\textbf{Transfer-}} & \multicolumn{2}{c}{\textbf{Optimization-}} \\ \cmidrule(l){4-7} 
         &  &  & \textbf{FGSM} & \textbf{PGD10} & \multicolumn{1}{c}{\textbf{\# Q}} & \textbf{FR} \\ \midrule \midrule
        \multirow{3.4}{*}{\rotatebox[origin=c]{90}{\scriptsize \textbf{CIFAR10}}} 
         & Baseline & 91\% & 1.00 & 1.00 & 13.7 & 100\% \\ \cmidrule(l){2-7} 
         & FF & 68\% & 0.38 & 0.58 & 901.8 & 80\% \\
         & \textbf{FL} & 91\% & \textbf{0.13} & \textbf{0.31} & \textbf{1131.9}\enspace\enspace\enspace & \textbf{74\%} \\ \midrule
        %
        \multirow{3.4}{*}{\rotatebox[origin=c]{90}{\scriptsize \textbf{ImageNet}}} 
         & Basline & 76\% & 1.00 & 1.00 & 13 & 100\% \\ \cmidrule(l){2-7} 
         & FF & 75\% & 0.51 & 0.93 & 140.7 & 99\% \\
         & \textbf{FL} & 73\% & \textbf{0.43} & \textbf{0.91} & \textbf{240.00} & \textbf{97\%} \\ \bottomrule
        %
    \end{tabular}
}
\caption{\textbf{Effectiveness of similarity-unparing in quantization.} We show the vulnerability of $f_s$ to black-box adversarial examples when we create quantized on-device models $f_{c'}$ by using three strategies. We highlighted the most effective defense strategy and the vulnerability it causes in bold.}
\label{tbl:defense-quantization}
\end{table}


\topic{Results.}
Table~\ref{tbl:defense-quantization} shows our results.
The baseline is the case where we quantize the pre-trained ResNet50 as-is.
We first find that FL is the best strategy at reducing the vulnerability to black-box adversarial examples.
In both CIFAR10 and ImageNet, the RFR of the transfer-based attack is reduced to 0.13--0.43 (FGSM) and 0.31--0.91 (PGD-10), compared to the baseline of 1.00.
Our defense 
increases the query complexity from 12--14 to 138--1132 and decreases the FR from 100\% up to 74\%.
However, we also find that FF is not an effective strategy for reducing vulnerability.
While it reduces the RFR in the transfer-based attacks, and the query efficiency and the FR in the optimization-based attack, the defense entails a large accuracy drop (91$\rightarrow$68\%).
In JF, we observe that our QAT results in a model that shows a significant accuracy drop when quantized (91$\rightarrow$47\% in CIFAR10).
We do not measure the vulnerability that $f_{c'}$ causes as our QAT 
leads to inaccurate models.
We analyze the reason for this failure.
We find that making a single model with 
significantly different gradients on the same inputs in its 32-bit and 8-bit representations is extremely difficult while achieving a high accuracy.

%
\subsection{Similarity Unpairing with Pruning}
\label{subsec:defense-pruning}

\noindent \textbf{Methodology.}
Pruning with FF is simple: we fine-tune the ResNet50 with our unpairing loss and then increase sparsity until the accuracy begins to drop, as we did before.
%
In FL, we first prune the pre-trained ResNet50 with 50\% sparsity and then fine-tune this model for 10--30 epochs with our unpairing objective.
%
%
%
Joint finetuning is again most difficult; we attempted to apply Variational Dropout~\cite{VariationalDropout} before the classification head of a network and use it a learned mask that removes neurons from the network.
During fine-tuning, we optimize only the dropout layer to minimize our similarity-unpairing objective.
Unfortunately, joint finetuning in this way does not work and reduces model accuracy to $20\%$.


%
\begin{table}[ht]
\centering
\adjustbox{max width=\linewidth}{
    \begin{tabular}{@{}ccccccc@{}}
        \toprule
        \multirow{2}{*}{\textbf{Dataset}} & \multirow{2}{*}{\textbf{Strategy}} & \multirow{2}{*}{\textbf{\begin{tabular}[c]{@{}c@{}}Acc.\\ (Pruned)\end{tabular}}} & \multicolumn{2}{c}{\textbf{Transfer-}} & \multicolumn{2}{c}{\textbf{Optimization-}} \\ \cmidrule(l){4-7} 
         &  &  & \textbf{FGSM} & \textbf{PGD10} & \textbf{\# Q} & \textbf{FR} \\ \midrule \midrule
        \multirow{3.5}{*}{\rotatebox[origin=c]{90}{\scriptsize \textbf{CIFAR10}}} 
         & Baseline & 88\% (0.5) & 0.89 & 1.00 & 19.5 & 100\% \\ \cmidrule(l){2-7} 
         & FF & 89\% (0.3) & 0.37 & 0.55 & 743.6 & 83\% \\
         & \textbf{FL} & 91\% (0.5) & \textbf{0.31} & \textbf{0.46} & \textbf{880.6} & \textbf{81\%} \\ \midrule
        %
        \multirow{3.5}{*}{\rotatebox[origin=c]{90}{\scriptsize \textbf{ImageNet}}} 
         & Basline & 73\% & 0.97 & 1.00 & 12.2 & 100\% \\ \cmidrule(l){2-7} 
         & \textbf{FF} & 73\% (0.4) & \textbf{0.51} & \textbf{0.94} & \textbf{148.18} & \textbf{99\%} \\
         & FL & 74\% (0.5) & 0.58 & 0.99 & 64.0 & 98\% \\ \bottomrule
        %
    \end{tabular}
}
\caption{\textbf{Effectiveness of similarity-unparing in pruning.} We show the vulnerability of $f_s$ to black-box adversarial examples when we use pruning to construct on-device models $f_{c'}$. The number in parenthesis next to each accuracy is the sparsity. We highlighted the most effective strategy in bold.}
\label{tbl:defense-pruning}
\end{table}


\noindent \textbf{Results.}
Table~\ref{tbl:defense-pruning} shows our results.
Pruning the pre-trained ResNet50 is our baselines.
%
Overall, the RFR of the transfer-based attacks is reduced to 0.31--0.51 (FGSM) and up to 0.46 (PGD-10) from 1.00.
In CIFAR10, our defense significantly increases the query complexity from 12--19 to 50--881 and decreases the FR from 100\% to 81\%.
We decrease the vulnerability similar to the case of retraining a model from scratch with 85--95\% \emph{fewer} training epochs.
In ImageNet, we reduce the RFR from 0.97$\rightarrow$0.51 (FGSM) and 1.00$\rightarrow$0.94 (PGD-10) while increasing query complexity by 10$\times$.
We note that this does not mean our defense, on ImageNet models, is less effective than retraining a model from scratch (a new $f_{\theta}$).
Our further analysis in Appendix~\ref{appendix:imagenet-finetune} implies that one could reduce the vulnerability more by increasing the learning rate or training with longer epochs.
In FF, we need to stop pruning at a smaller sparsity (0.3--0.4).
Pruning more than 30--40\% of the parameters leads to an accuracy drop of more than 20\%.
Unfortunately, in JF, our proposed pruning makes the accuracy of a fine-tuned model to $\sim$20\%.
%
%

%
\subsection{Similarity Unpairing with Distillation}
\label{subsec:defense-distillation}

\noindent \textbf{Methodology.}
We further examine the three strategies 
to achieve similarity-unpairing during the distillation process.
We can first fine-tune the teacher with our unpairing objective 
and use the fine-tuned teacher for distillation (FF).
Here, we hypothesize that the dissimilarity we encode to the teacher can be transferred to the student.
We can also fine-tune the student model for a few epochs after the distillation, \textit{i.e.}, we directly reduce the vulnerability.
Moreover, we can use our objective function as a regularizer for the distillation loss (JF).
In this case, we only consider the input-gradient similarity 
as minimizing the output similarity can disturb the distillation process, 
\textit{i.e.}, the distillation does not work 
if we make the outputs of both the teacher and student dissimilar.


%
\begin{table}[ht]
\centering
\adjustbox{max width=\linewidth}{
    \begin{tabular}{@{}cccccrc@{}}
        \toprule
        \multirow{2}{*}{\textbf{\begin{tabular}[c]{@{}c@{}}Network\\ (Student)\end{tabular}}} & \multirow{2}{*}{\textbf{Strategy}} & \multirow{2}{*}{\textbf{\begin{tabular}[c]{@{}c@{}}Acc.\\ (Student)\end{tabular}}} & \multicolumn{2}{c}{\textbf{Transfer-}} & \multicolumn{2}{c}{\textbf{Optimization-}} \\ \cmidrule(l){4-7} 
         &  &  & \textbf{FGSM} & \textbf{PGD10} & \textbf{\# Q} & \textbf{FR} \\ \midrule \midrule
        \multirow{5}{*}{\rotatebox[origin=c]{90}{\textbf{ResNet18}}} 
         & Baseline & 88\% & 0.18 & 0.70 & 752.3 & 85\% \\ \cmidrule(l){2-7} 
         & FF & 90\% & 0.19 & 0.73 & 413.5 & 94\% \\
         & \textbf{FL} & 84\% & \textbf{0.12} & \textbf{0.37} & \textbf{2337.1} & \textbf{43\%} \\ 
         & JF & 89\% & 0.22 & 0.85 & 352.6 & 94\% \\ \midrule
        \multirow{4}{*}{\rotatebox[origin=c]{90}{\textbf{NASNet}}} 
         & Baseline & 82\% & 0.09 & 0.40 & 1713.7 & 62\% \\ \cmidrule(l){2-7} 
         & FF & 87\% & 0.11 & 0.52 & 1052.4 & 79\% \\
         & \textbf{FL} & 88\% & \textbf{0.09} & \textbf{0.41} & \textbf{1114.2} & \textbf{78\%} \\ 
         & JF & 89\% & 0.15 & 0.69 & 722.1 & 86\% \\ \bottomrule
    \end{tabular}
}
\caption{\textbf{Effectiveness of similarity-unparing in distillation.} We analyze the effectiveness of our defense in distillation. We use two students ResNet18 and NASNet in CIFAR10. We highlighted the most effective cases in bold.}
\label{tbl:defense-distillation}
\end{table}


\noindent \textbf{Results.}
Table~\ref{tbl:defense-distillation} shows our results.
Our objective function can reduce vulnerability in all three cases.
We run our experiments in CIFAR10 
and consider two student networks, ResNet18 and NASNet.
We first observe that both FL and JF achieve the least vulnerability to black-box adversarial examples.
Compared to the baseline, 
the RFR reduces by $\sim$0.10 in FGSM and 0.3--0.4 in PGD-10.
The query complexity increases by 2--7$\times$, 
and the FR is reduced by 10--50\%.
We also find that FF is more effective in distillation than quantization or pruning.
Distillation makes the outputs of a student resemble those of a teacher---the fine-tuned model in our case.
Our analysis further shows that 
one can jointly optimize both the distillation and similarity-unpairing objectives.
Unlike the JF strategy used in quantization and pruning, 
this can reduce the vulnerability while preserving the accuracy of a student.

We additionally compare our similarity-unpairing 
with the adversarially-robust distillation mechanism proposed by Goldblum~\textit{et al.}~\cite{goldblum2020adversarially}.
We take the ResNet50 model as a teacher and perform this distillation process onto ResNet18 (a student).
The student achieves an accuracy of 72\%,
which is 16\% less than the models constructed by our defense mechanism.
In the transfer-based attacks, 
we observe the RFR of 0.04 and 0.26 in FGSM and PGD-10, respectively.
The optimization-based attacks achieve the FR of 44\% 
with 2398.3 queries per adversarial-example crafting on average.
The robust distillation offers a similar amount of vulnerability decrease, 
but it sacrifices the accuracy of the resulting models by 10--15\%.
%

%
\subsection{Similarity Unpairing with New Architectures}
\label{subsec:defense-new-architectures}


\noindent \textbf{Methodology.}
Here, we examine the scenarios 
where a victim chooses a compact architecture, 
designed by architecture search mechanisms (\textit{e.g.}, NASNet) or manually (\textit{e.g.}, MobileNet or SqueezeNet).
In this case, we consider the FL or JF strategies 
as the victim needs to train a network from scratch.
We wonder if the defender can push the vulnerability reduction 
further than just training a new architecture from scratch.
In FL, we take a pre-trained model 
and fine-tune it with our similarity-unpairing objective for 40 more epochs.
In JF, we train a network from scratch with our defense objective 
using the same hyper-parameter for training the pre-trained models.


%
\begin{table}[h]
\centering
\adjustbox{max width=\linewidth}{
    \begin{tabular}{@{}cccccrc@{}}
        \toprule
        \multirow{2}{*}{\textbf{Dataset}} & \multirow{2}{*}{\textbf{Net-arch.}} & \multirow{2}{*}{\textbf{Acc.}} & \multicolumn{2}{c}{\textbf{Transfer-}} & \multicolumn{2}{c}{\textbf{Optimization-}} \\ \cmidrule(l){4-7} 
         &  &  & \textbf{FGSM} & \textbf{PGD10} & \textbf{\# Q} & \textbf{FR} \\ \midrule \midrule
        \multirow{4}{*}{\rotatebox[origin=c]{90}{\textbf{CIFAR10}}} 
         & \multirow{2}{*}{MobileNetV2} & 93\% & 0.24 & 0.76 & 580.3 & 89\% \\
         & & 92\% & 0.12 & 0.35 & 1173.4 & 75\% \\ \cmidrule(l){2-7} 
         & \multirow{2}{*}{SqueezeNet} & 89\% & 0.09 & 0.47 & 1547.1 & 67\% \\ 
         & & 87\% & 0.09 & 0.43 & 1345.2 & 73\% \\ \bottomrule
        %
        %
    \end{tabular}
}
\caption{\textbf{Effectiveness of similarity-unpairing in using different architectures.} We analyze the effectiveness of our defense in fine-tuning on-device models with smaller architectures (\textit{i.e.}, MobileNetV2 and SqueezeNet). We only consider the finetune-last strategy. For each architecture, the first row is the baseline, and the second row is our results.}
\label{tbl:defense-newarch}
\end{table}

\noindent \textbf{Results.}
Table~\ref{tbl:defense-newarch} shows our results.
We run our experiments with MobileNet and SqueezeNet in CIFAR10.
We only report the results from FL as 
we find JF training has a 20--30\% accuracy drop. 
In FL, we find that our defense reduces the vulnerability to transfer-based attacks by 50\% at most.
Compared to the baseline where we use undefended MobileNetV2,
the fine-tuned model shows 0.12--0.36 less in RFR.
In the optimization-based attack, 
the query efficiency and the FR are reduced by 2$\times$ and 14\%.
In SqueezeNet, both the undefended and fine-tuned models show a similar vulnerability.
SqeezeNet already achieves low vulnerability 
even if the attacker uses the undefended model; 
thus, we marginally improves the security.
%

%

%
\subsection{Similarity Unpairing with Once-for-All}
\label{subsec:defense-ofa}

\noindent \textbf{Methodology.}
We finally examine our unpairing's effectiveness 
when used with the once-for-all (OFA) paradigm~\cite{OFA}.
We only consider the FL strategy.
FF is not compatible with this paradigm 
as it derives multiple sub-networks from a pre-trained model%
---the vulnerability between the pre-trained model and sub-networks will remain the same.
We examine JL, but the computations require to optimize our objective 
while we run the progressive shrinking algorithm is intractable.
We consider the OFANet-267---the most vulnerable case---and fine-tune it for 20 epochs with our defense. 

\topic{Results.}
We find that the fine-tuning reduces the vulnerability of the server model (OFANet-467) to black-box adversarial examples 
while maintaining a 74\% accuracy.
In the transfer-based attacks, 
we achieve a RFR of 0.65 and 0.98 in FGSM and PGD-10, 
compared to the baseline of 0.94 and 1.00.
In the optimization-based attack, 
the query complexity increases from 28 to 220 on average, 
but the FR remains almost the same (100\%$\rightarrow$97\%).
%
%
We achieve this improvement
even when the sub-networks are only fine-tuned for a few epochs.
%


\section{Conclusion}
\label{sec:conclusion}

This paper introduces a new security consideration when training machine learning models
that will be hosted both server-side but also published on-device:
the release of the on-device model should not increase the server-side vulnerability to adversarial examples.
We have shown that naive release methods, as studied in efficient deep learning, 
do significantly increase server-side
vulnerability, but that by fine-tuning an efficient model before the release 
with our \emph{similarity-unpairing}, it is possible to reduce the advantage an adversary would have from using this on-device model significantly.
We make two categories of conclusions:

\topic{Next steps for researchers.}
We have posed a new research problem, 
and while we believe in having constructed
a robust defense, no defense can be perfect.
It is an open question if there is a \emph{new} way to design 
stronger black-box attacks that reduce the effectiveness of similarity-unpairing.

We have answered one question in this direction, but other questions remain.
For simplicity, we have studied the case where the defender releases just an efficient model;
in practice, however, model providers often update their model over time to
improve accuracy.
It is unknown whether releasing multiple models with small parameter differences 
can increase adversarial vulnerability even further
or if this practice results in a break of our defense without stronger attacks.

We have shown a trade-off between 
the costs required for constructing on-device models and security.
The question still remains unanswered 
if this trade-off is inherent to the problem or if there is a way to design around it.

\topic{Next steps for practitioners.}
The security threat we expose is immediately practical 
in any setting where both server- and on-device
models are constructed.
If security is a potential consideration,
this vulnerability will need to be considered when
evaluating the vulnerability of the server-side model.

One immediate consequence of
this threat is that any server model where corresponding efficient models
have \emph{already} been released should be considered insecure.
Future on-device models should only be released if either security is not a consideration
or if steps have been taken to reduce the vulnerability.
Fortunately, 
we have found with our defense that even a small amount of fine-tuning
can mitigate adversaries from exploiting the on-device model to launch adversarial attacks on the server model more successfully.

We believe that answering those open questions will bring the two seemingly distant objectives closer: 
improving the security of server-side models against black-box attacks and releasing computationally-efficient on-device models.




\section*{Acknowledgment}
\label{sec:ack}

We thank anonymous reviewers for constructive feedback.


\bibliographystyle{IEEEtran}
\bibliography{bib/security}


\appendices

\section{Practicality of the Threat Model}
\label{appendix:practical-attack-scenarios}

We discuss practical scenarios
where an adversary exploits on-device models 
to attack original, server-side models.

\topic{A server-side model for many services.}
The victim can train a (server-side) model
and uses it as a building block for multiple services they offer.
For example, service providers like Google
can train a classifier to detect NSFW photos and uses it
(1) to filter out the images uploaded to Google Drive and
(2) to detect YouTube videos with sensitive content. 
They can also train a model 
that acts as a profanity filter and uses it
(1) to filter out spam emails and 
(2) to remove websites containing bad language from their search results.

\topic{Evading those filters is challenging.}
As the models (i.e., filters) are typically deployed to the servers,
where an adversary does not have white-box access,
they harness black-box attacks to evade the filtering.
However, black-box attacks, such as transfer-based or query-based attacks,
are less successful or require a high query complexity
than the white-box attacks.
The victim can also deploy security mechanisms,
such as rate-limiting, user authentication, or returning only hard labels,
that increases the costs of black-box attacks.

\topic{Adversaries can exploit on-device models for evasion.}
To overcome this challenge, 
the attacker can use on-device models the victim releases.
Suppose that a company pushes a NSFW filter to mobile devices
to reduce the networking bandwidth and protect user privacy.
As the filter will be available to anyone has a mobile device,
the attacker reverse-engineer the filter and
use it to generate adversarial examples and test their attack.
Since there is no limit on what the adversary can do with the device,
it is fair to assume the model will be on the attacker's hand.
Once the model is there,
the attacker can use it for generating videos
that evades the company's filtering mechanisms.
It will be the same for the profanity filter case.

\section{Experimental Setup in Detail}
\label{appendix:experimental-details}

\noindent \textbf{Setup.}
We implemented our analysis framework using Python v3.8 
and PyTorch v1.8.0\footnote{\url{https://pytorch.org/}} 
that supports CUDA 11.3 for accelerating computations by GPUs.
We run all our experiments on a computing cluster, where we have a machine equipped with 2 Intel Xeon 6248R 3.00GHz 48-core processors, 512GB of RAM, and 8 Nvidia Tesla V100 and 8 Nvidia A40 GPUs.

\smallskip
\noindent \textbf{Hyperparameters.}
We train our CIFAR10 network for 200 epochs from scratch 
using the same set of hyper-parameters following the original study~\cite{ResNet}.
We use pre-trained ImageNet models offered by Torchvision library\footnote{\url{https://pytorch.org/vision/stable/models.html}}.
In the case of training these models from scratch, we use the same training configurations used for constructing the pre-trained models\footnote{\url{https://github.com/pytorch/vision/tree/main/references/classification}}.

\section{Discussion about the Metrics\\for Quantifying the Vulnerability}
\label{appendix:other-metrics}

\topic{Setup.}
We evaluate the effectiveness of the metrics,
widely-used in the field of adversarial examples,
in capturing the adversarial vulnerability in our setting.
We examine two popular metrics:
(i) the accuracy drop of a model---typically used in the prior work on adversarial examples, and 
(ii) the metrics proposed in a recent work~\cite{Transfer:ICML'21} on transferability of adversarial examples.
We use them to quantify the vulnerability in Table~\ref{tbl:effectiveness-of-priors}.


\begin{table}[ht]
\centering
\adjustbox{max width=\linewidth}{
    \begin{tabular}{@{}cc|lcc|cc|cc|cc@{}}
    \toprule
    \multicolumn{2}{c|}{\textbf{Server $f_s$}} & \multicolumn{3}{c|}{\textbf{On-device $f_o$}} & \multicolumn{2}{c|}{\textbf{Metrics in~\cite{Transfer:ICML'21}}} & \multicolumn{4}{c}{\textbf{Acc. drop (on $f_s$ and $f_o$)}} \\ \midrule
    \textbf{Arch.} & \textbf{Acc.} & \textbf{Mech.} & \textbf{Arch} & \textbf{Acc.} & \textbf{FGSM} & \textbf{PGD10} & \multicolumn{2}{c|}{\textbf{FGSM}} & \multicolumn{2}{c}{\textbf{PGD10}} \\ \midrule \midrule
    \multirow{7}{*}{\rotatebox[origin=c]{90}{\textbf{R50}}}%
     & \multirow{7}{*}{92\%} & as-is & R50 & 92\% & 1.00 & 1.00 & 61\% & 61\% & 92\% & 92\% \\
     &  & new $f_c$ & R50 & 91\% & 0.28 & 0.22 & 74\% & 16\% & 91\% & 79\% \\ \cmidrule{3-11}
     &  & Quant & R50 & 92\% & 1.00 & 1.00 & 60\% & 60\% & 92\% & 92\% \\
     &  & Pruning & R50 & 91\% & 0.97 & 0.97 & 61\% & 58\% & 91\% & 91\% \\
     &  & Distill & R18 & 89\% & 0.26 & 0.18 & 59\% & 12\% & 89\% & 70\% \\
     &  & Manual & MV2 & 93\% & 0.27 & 0.16 & 68\% & 16\% & 93\% & 67\% \\
     &  & Manual & SN & 89\% & 0.12 & 0.09 & 59\% & 4\% & 89\% & 40\% \\ \bottomrule
    \end{tabular}
}
\caption{\textbf{Comparing the metrics proposed by prior work.} We evaluate the two vulnerability metrics offered in literature: the accuracy drop caused by adversarial examples and the metrics proposed by~\cite{Transfer:ICML'21} in CIFAR10. We compute the metrics between the pair of server and on-device models we examined in Table~\ref{tbl:effectiveness-of-priors}. We use adversarial examples crafted by FGSM and PGD-10 on the on-device models.}
\label{tbl:compare-metrics}
\end{table}


\topic{Results.}
Table~\ref{tbl:compare-metrics} shows our results.
R50, R18, MV2, and SN refer to ResNet50, ResNet18, MobileNetV2, and SqueezeNet architectures, respectively.
In each column from the left, 
where we measure the accuracy drop, 
we show the accuracy degradation of on-device $f_{\theta_o}$ and server $f_{\theta_s}$ models 
caused by adversarial examples crafted on $f_{\theta_o}$.
We first observe that within each attack case, 
the metric captures the vulnerability somehow, 
i.e., if the vulnerability decreases, the metrics are reduced as well.
However, we also find that those metrics are sometimes misleading.
For example, the metrics~\cite{Transfer:ICML'21} 
are smaller in PGD-10 (0.18) than in FGSM (0.26), 
while PGD-10 is a much stronger attack.
In distillation, 
the accuracy drop of the server model caused by PGD-10 (70\%) 
is much larger than that of FGSM (12\%).

While the accuracy drop (or the fooling rate) 
captures the strength of an attack well, 
we observe that it cannot capture 
how many adversarial examples successful on $f_{\theta_o}$ can also fool the attack target ($f_{\theta_s}$).
For example, the adversary would not use a set of adversarial examples, 
unsuccessful on $f_{\theta_o}$---the model used to craft them---for attacking the server-side model.
To address this problem, 
we propose a new metric, relative fooling rate (RFR), 
that can capture how much the attacker can be successful relatively more/less than the baseline.
In our work, we set the baseline as the white-box attacks on $f_{\theta_s}$ and 
compare them with the black-box attacks with $f_{\theta_o}$.

\section{Computing Similarities}
\label{appendix:sim-metrics}

In \S\ref{subsec:similarity-between-the-two}, we compute our similarity metrics as follows:
\begin{itemize}[leftmargin=1.2em, noitemsep, topsep=0.2em]
    \item \textbf{Output} similarity defined as:
    \begin{align*}
        {C_s}^{\text{o}} = \text{mean}_{\mathcal{D}_{\text{s}}} \big\{ cosine(f_{\theta_s}(x), f_{\theta_o}(x)) \big\}
    \end{align*}
    where $f_{\theta_s}$ and $f_{\theta_o}$ are server-side and on-device models;
    $(x, y)$ is an input sample drawn from 1,000 randomly chosen training samples $\mathcal{D}_{\text{s}}$;
    $f(x)$ is the logits;
    and ${C_s}^{\text{o}}$ is the cosine similarity.
    We compute the mean over $\mathcal{D}_{\text{s}}$.
    %
    \item \textbf{Input-gradients} similarity defined as:
    \begin{align*}
        {C_s}^{\text{i}} = \text{mean}_{\mathcal{D}_{\text{s}}} 
            \big\{ cosine( \nabla_{\text{x}} \mathcal{L}(f_{\theta_s}, x, y), \nabla_{\text{x}} \mathcal{L}(f_{\theta_o}, x, y)) \big\}
    \end{align*}
    where $\mathcal{L}$ is the loss function (cross-entropy);
    $\nabla_{\text{x}} \mathcal{L}(f, x, y)$ is the gradient of the loss with respect to an input $x$.
    The others are the same as the output-level similarity.
    %
    \item \textbf{Activation} similarity defined as:
    \begin{align*}
        {C_s}^{\text{a}} = \text{mean}_{\mathcal{D}_{\text{s}}} 
            \big\{ cosine( z_{\text{s}}(x), z_{\text{o}}(x) ) \big\}
    \end{align*}
    where $z_{\text{s}}(x)$ is the activtaions of an input $x$ computed by $f_{\theta_s}$.
    Here, we use the penultimate layer's activations for $z$.
\end{itemize}

\section{Limitations of Using Activation Similarity}
\label{appendix:limit-in-activation-sim}

In this section, 
we discuss more why activation-space similarity 
is not desirable in measuring adversarial vulnerability.
As we expected, activation similarity is much more fragile.
First, note we can only compare activation vectors 
when they share the same dimensionality, 
e.g., ResNets that have 512-dimensional latent representations.
But also, when training from scratch, 
the activation vectors can easily represent the same feature 
but be ``permuted'' and thus show a near-zero similarity 
even if they represent the same data.
And this is not just something that is technically possible, 
it actually does easily happen when we train models separately from scratch.

\begin{figure}[ht]
\centering
\includegraphics[width=0.8\linewidth]{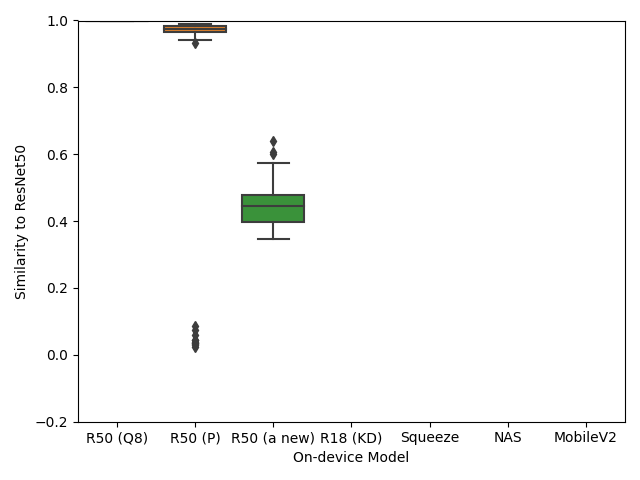}
\caption{\textbf{Similarity metrics measured in the latent representation space.} We measure the similarities between the server (ResNet50) and on-device models in Table~\ref{tbl:effectiveness-of-priors}. We can compare them between models with the same latent dimension; otherwise, we cannot compute this metric.}
\label{fig:activation-similarity}
\end{figure}


\begin{table*}[t]
\centering
\adjustbox{max width=\textwidth}{
    \begin{tabular}{@{}c|cccrr|cccrr|cccrr@{}}
    \toprule
    \textbf{Objective} & \multicolumn{5}{c|}{\textbf{Penalize Output-level Sim.}} & \multicolumn{5}{c|}{\textbf{Penalize Feature-level Sim.}} & \multicolumn{5}{c}{\textbf{Penalize Input-level Sim.}} \\ \midrule
    \multirow{2}{*}{$\lambda$} & \textbf{-} & \multicolumn{2}{c}{\textbf{Transfer-}} & \multicolumn{2}{c|}{\textbf{Optimization-}} & \textbf{-} & \multicolumn{2}{c}{\textbf{Transfer-}} & \multicolumn{2}{c|}{\textbf{Optimization-}} & \textbf{-} & \multicolumn{2}{c}{\textbf{Transfer-}} & \multicolumn{2}{c}{\textbf{Optimization-}} \\ \cmidrule{3-6} \cmidrule{8-11} \cmidrule{13-16}
     & \textbf{Acc.} & \textbf{FGSM} & \textbf{PGD10} & \textbf{\# Q} & \textbf{FR} & \textbf{Acc.} & \textbf{FGSM} & \textbf{PGD10} & \textbf{\# Q} & \textbf{FR} & \textbf{Acc.} & \textbf{FGSM} & \textbf{PGD10} & \textbf{\# Q} & \textbf{FR} \\ \midrule \midrule
    0.0 & 92\% & 0.25 & 0.91 & 274.4 & 95\% & 92\% & 0.25 & 0.91 & 274.4 & 95\% & 92\% & 0.25 & 0.91 & 274.4 & 95\% \\ \midrule
    0.001 & 80\% & 0.05 & 0.23 & 2979.0 & 48\% & 80\% & 0.04 & 0.23 & 2410.0 & 46\% & 93\% & 0.25 & 0.88 & 344.1 & 94\% \\
    0.01 & 81\% & 0.04 & 0.23 & 2196.0 & 51\% & 81\% & 0.07 & 0.41 & 1775.5 & 62\% & 91\% & 0.25 & 0.88 & 347.0 & 94\% \\
    0.1 & 82\% & 0.04 & 0.30 & 1903.0 & 58\% & 77\% & 0.02 & 0.10 & 2898.3 & 31\% & 91\% & 0.25 & 0.81 & 396.0 & 93\% \\
    1.0 & 81\% & 0.02 & 0.14 & 2711.8 & 37\% & 80\% & 0.04 & 0.20 & 2521.4 & 43\% & \textbf{90\%} & \textbf{0.12} & \textbf{0.55} & \textbf{1237.0} & \textbf{75\%} \\
    10.0 & 85\% & 0.10 & 0.56 & \multicolumn{1}{c}{-} & \multicolumn{1}{c|}{-} & 79\% & 0.05 & 0.30 & \multicolumn{1}{c}{-} & \multicolumn{1}{c|}{-} & 89\% & 0.02 & 0.04 & \multicolumn{1}{c}{-} & \multicolumn{1}{c}{-} \\ \bottomrule
    \end{tabular}
}
\caption{\textbf{Effectiveness of our similarity-unpairing defense (train from scratch).} We measure the vulnerability of $f_s$ to black-box adversarial examples when the defender trains $f_{c}$ with our defense objective instead of doing fine-tuning. The settings are inherited from our experiments in Table~\ref{tbl:effectiveness-of-unpairing}. We make the cases where the attacker is the least successful in bold.}
\label{tbl:effectiveness-of-unpairing-scratch}
\end{table*}


\topic{Dimensions should be the same.}
Our activation-space similarity predicts the vulnerability between the models when the dimensions of their latent representations are the same.
In Table~\ref{fig:activation-similarity}, in ResNets, the vulnerability reduces as the similarity decreases.
However, not all the models have the same dimensions there.
For example, while ResNets in CIFAR10 typically use a 256-dimensional vector in their latent representations, i.e., the activations just before the classification head, MobileNetV2 uses 1024-dimensions.
Thus, we cannot compute the similarity loss when latent dimensions are different.

\topic{Models that share the same architecture can learn different features.}
Another assumption in measuring the activation similarity is that two models using the same architecture, trained individually, will learn similar representations.
However, we can also hypothesize that there could be a permutation of a similar set of representations.
Suppose that a representation vector $[l_1, l_2, ..., l_n]$ for an input observed from one network, the second network can have $[l_n, l_1, ..., l_2]$ where elements are permuted.
In this case, the vector contains the same set of elements, but the cosine similarity between the two cannot be 1.0 unless all the $l_i$'s are the same.
We leave the further investigation of this issue as future work.

\section{Similarity Reduction in Training from Scratch}
\label{appendix:additional-results-sim-reduction}

\noindent \textbf{Single objective at a time.}
In Table~\ref{tbl:effectiveness-of-unpairing-scratch}, we examine whether we can achieve a better reduction in vulnerability when we train on-device models from scratch.
We hypothesize that training from scratch may help the optimization process of our objective by increasing the search space in the loss surface.
In contrast, fine-tuning has a limited search space for an optimum as the process only explores a smaller region around the location where a pre-trained model is.
We train each on-device model from scratch using the same hyper-parameters we use to construct pre-trained models.
We vary the hyper-parameter $\lambda_{i}$ in 0.001--10.0 to control the impact of our unpairing loss term.
In each case, we measure the accuracy of the fine-tuned model $f_{\theta_{c'}}$ and the vulnerability metrics we define.
As a baseline, we set $\lambda_{i}$ to zero and include the result in the first row.

Interestingly, we find that training from scratch does not offer a reduction of the vulnerability increase.
The attacker is the least successful when penalizing the input-gradients similarity while training an on-device model from scratch.
The model achieves an accuracy of 90\% while reducing the RFR by 0.13 (FGSM) and by 0.36 (PGD-10).
In optimization-based attacks, the query complexity increases from 274 to 1237, and the FR decreases by 20\%.
However, penalizing the output-level or feature-level similarities leads to client models with 7--15\% less accuracy.
We observe the reduced vulnerability in those cases, but the victory may be useless as a defender ends up deploying these inaccurate models to devices.


\begin{table}[h]
\centering
\adjustbox{max width=\linewidth}{
    \begin{tabular}{@{}cc|c|cc|rc@{}}
    \toprule
    \multicolumn{2}{c|}{\textbf{Objectives}} & \multicolumn{5}{c}{\textbf{Penalize Multiple Sim.}} \\ \midrule
    \textbf{Output.} & \textbf{Input.} & \textbf{-} & \multicolumn{2}{c|}{\textbf{Transfer-based}} & \multicolumn{2}{c}{\textbf{Optimization-based}} \\ \cmidrule{3-7}
    \textbf{$\lambda_1$} & \textbf{$\lambda_2$} & \textbf{Acc.} & \textbf{FGSM} & \textbf{PGD10} & \multicolumn{1}{c}{\textbf{\#Q}} & \textbf{FR} \\ \midrule \midrule
    0.01 & 0.01 & 90\% & 0.82 & 1.00 & 50.7 & 99\% \\
    0.01 & 0.1 & 92\% & 0.84 & 1.00 & 132.2 & 99\% \\
    \textbf{0.01} & \textbf{1.0} & \textbf{90}\% & \textbf{0.07} & \textbf{0.06} & \textbf{3492.0} & \textbf{14}\% \\ \midrule
    0.1 & 0.01 & 92\% & 0.56 & 0.73 & 551.6 & 89\% \\
    0.1 & 0.1 & 91\% & 0.39 & 0.74 & 588.2 & 89\% \\ 
    0.1 & 1.0 & 90\% & 0.08 & 0.07 & 3463.2 & 14\% \\ \midrule
    1.0 & 0.01 & 89\% & 0.69 & 0.72 & 674.98 & 85\% \\
    1.0 & 0.1 & 89\% & 0.61 & 0.70 & 635.61 & 86\% \\
    1.0 & 1.0 & 90\% & 0.34 & 0.51 & 824.9 & 82\% \\
    %
    \bottomrule
    \end{tabular}
}
\caption{\textbf{Combining multiple unpairing objectives (train from scratch).} We measure the vulnerability when we train a client model from scratch instead of doing fine-tuning. We penalize the outputs and input-gradients similarity. The least vulnerable case is highlighted in bold.}
\label{tbl:combine-multiple-objectives-scratch}
\end{table}



%
\begin{table*}[t]
\centering
\adjustbox{max width=\textwidth}{
    \begin{tabular}{c|lc|llcc|cc|rr}
    \toprule
    \multirow{2}{*}{\textbf{Dataset}} & \multicolumn{2}{c|}{\textbf{Original Model ($f_s$)}} & \multicolumn{4}{c|}{\textbf{On-device Model ($f_o$)}} & \multicolumn{2}{c|}{\textbf{Transfer-based ($\ell_{\inf}$)}} & \multicolumn{2}{c}{\textbf{Optimization-based}~\cite{P-RGF}} \\ \cmidrule{2-11}
     & \multicolumn{1}{c}{\textbf{Arch.}} & \textbf{Acc. (\%)} & \multicolumn{1}{c}{\textbf{Mechanism}} & \multicolumn{1}{c}{\textbf{Arch. (New)}} & \textbf{Train} & \textbf{Acc. (\%)} & \textbf{FGSM} & \textbf{PGD-10} & \textbf{\# Queries} & \textbf{Success (\%)} \\ \midrule \midrule
    \multirow{10}{*}{\rotatebox[origin=c]{90}{\textbf{CIFAR10}}} 
     & \multirow{10}{*}{ResNet50} & \multirow{10}{*}{\facc{80.2}{0.3}} & Baseline (as-is) & ResNet50 (\xmark) & \xmark & \facc{80.2}{0.3} & \facc{1.00}{0.00} & \facc{1.00}{0.00} & \facc{1476.1}{\hspace{0.5em}43.4} & \facc{64}{1.1} \\
     &  &  & Baseline (a new) & ResNet50 (\xmark) & \cmark & \facc{80.2}{0.3} & \facc{0.64}{0.07} & \facc{0.65}{0.02} & \facc{2075.3}{\hspace{0.5em}17.3} & \facc{49}{0.5} \\
     &  &  & Baseline (no $f$) & \multicolumn{1}{c}{-} & - & - & - & - & \facc{3105.5}{\hspace{0.5em}31.4} & \facc{23}{0.7} \\ \cmidrule{4-11} 
     &  &  & Quantization & ResNet50 (\xmark) & \xmark & \facc{80.2}{0.3} & \facc{1.00}{0.00} & \facc{1.00}{0.00} & \facc{1474.9}{\hspace{0.5em}43.7} & \facc{64}{1.1} \\
     &  &  & Pruning & ResNet50 (\xmark) & \xmark & \facc{76.8}{0.7} & \facc{0.77}{0.06} & \facc{0.80}{0.04} & \facc{1885.9}{116.7} & \facc{53}{3.0} \\
     &  &  & Distillation & ResNet18 (\cmark) & \cmark & \facc{87.7}{0.2} & \facc{0.05}{0.01} & \facc{0.05}{0.02} & \facc{3046.0}{\hspace{0.5em}33.8} & \facc{25}{0.8} \\
     &  &  & Distillation & NASNet (\cmark) & \cmark & \facc{81.0}{0.8} & \facc{0.07}{0.02} & \facc{0.10}{0.03} & \facc{3011.3}{\hspace{0.5em}46.5} & \facc{26}{1.2} \\
     &  &  & NAS & NASNet (\cmark) & \cmark & \facc{75.2}{0.5} & \facc{0.52}{0.04} & \facc{0.52}{0.04} & \facc{2301.8}{\hspace{0.5em}10.5}& \facc{43}{0.4} \\
     &  &  & Manual-arch.& MobileNetV2 (\cmark) & \cmark & \facc{80.8}{2.2} & \facc{0.61}{0.03} & \facc{0.61}{0.06} & \facc{2124.6}{\hspace{0.5em}63.1} & \facc{47}{1.6} \\
     &  &  & Manual-arch. & SqueezeNet (\cmark) & \cmark & \facc{78.2}{0.3} & \facc{0.62}{0.02} & \facc{0.62}{0.06} & \facc{2108.9}{\hspace{0.5em}32.4} & \facc{48}{0.8} \\ \bottomrule
    \end{tabular}
}
\caption{\textbf{Vulnerability to black-box attacks when on-device models are used as priors.}
Six different mechanisms are used for producing on-device models, 
and we also note whether or not each technique requires additional training.
All the on-device models are constructed from the robust ResNet50 trained with PGD-7.
FGSM, PGD-10, and P-RGF are used. 
We measure the relative fooling rate (RFR) for FGSM and PGD-10, and for P-RGF, we measure the \# of queries and the FR.}
\label{tbl:effectiveness-of-adv-priors}
\end{table*}



\topic{Combining multiple unpairing objectives.}
In Table~\ref{tbl:combine-multiple-objectives-scratch}, we observe that combining multiple objectives while training a model from scratch can effectively reduce the vulnerability.
We use the same settings as our experiments in \S\ref{subsec:defense-setup}.
The hyper-parameters $\lambda_1$ and $\lambda_2$ are the weights for the output similarity loss and input-gradient similarity loss.
We vary them from 0.01--1.0.
We exclude the activation-level similarity as the objective is ineffective.

By setting $\lambda_1$ and $\lambda_2$ to 0.01 and 1.0, we achieve the RFRs of 0.07 and 0.06 in FGSM and PGD-10.
We also increase the query complexity to 3492.0 and reduce the FR to 14\% in the optimization-based attacks.
The result is compatible with the best case we observe in the fine-tuning scenarios.

We further show that the final model's accuracy decreases as we increase the importance of the output similarity ($\lambda_{1}$).
Compared to the cases of setting $\lambda_{1}$ to 0.01, the final models trained with $\lambda_{1}=1.0$ have 2--3\% reduced accuracy.
It indicates that making the outputs from the two models \emph{dissimilar} can disturb the training process from finding an optimum.

\section{A Note on the ImageNet Fine-tuning}
\label{appendix:imagenet-finetune}


%
\begin{figure}[ht]
    \minipage{0.49\linewidth}
	  \includegraphics[width=\linewidth]{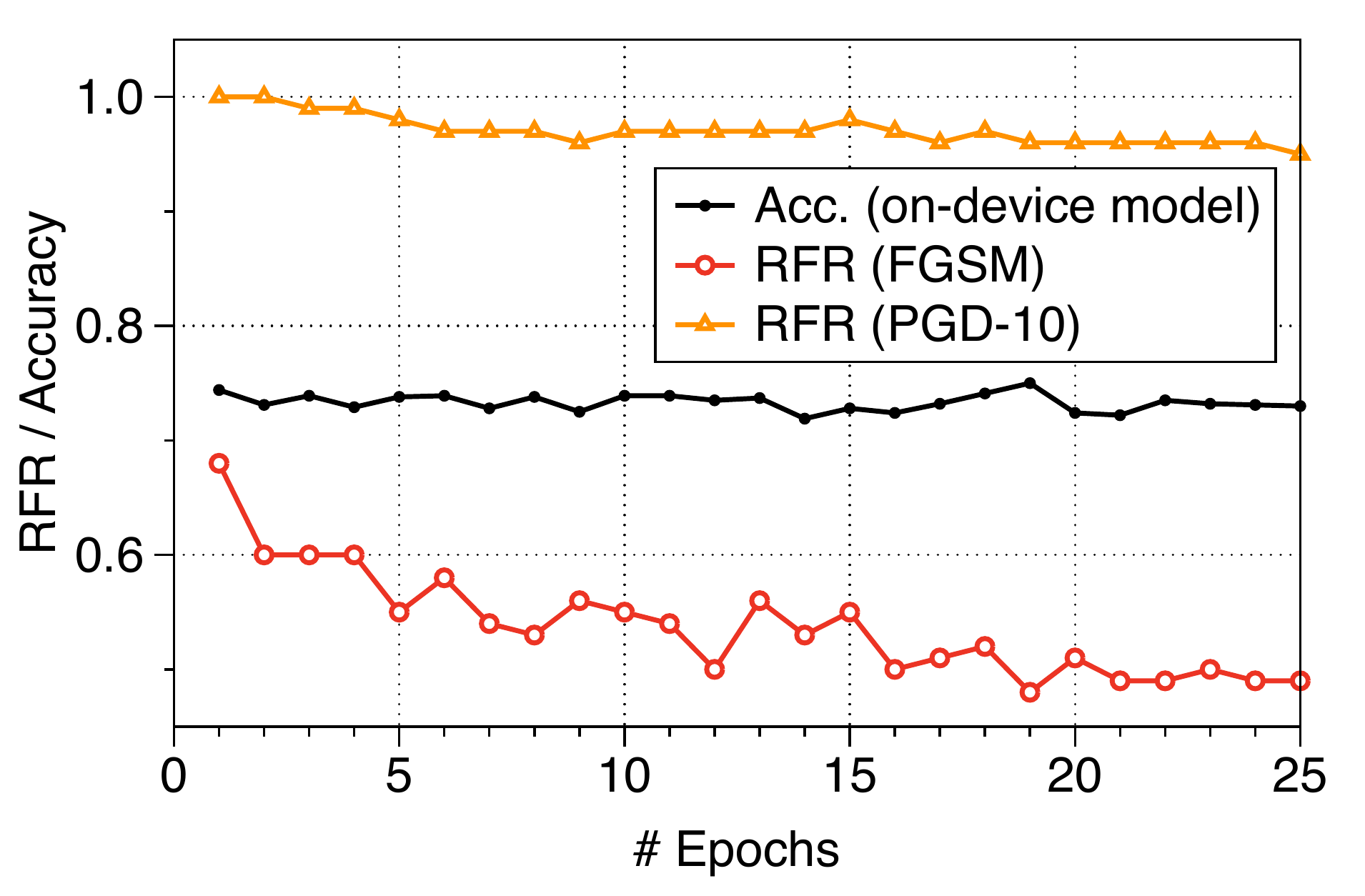}
    \endminipage
    \hfill
    \minipage{0.49\linewidth}
      \includegraphics[width=\linewidth]{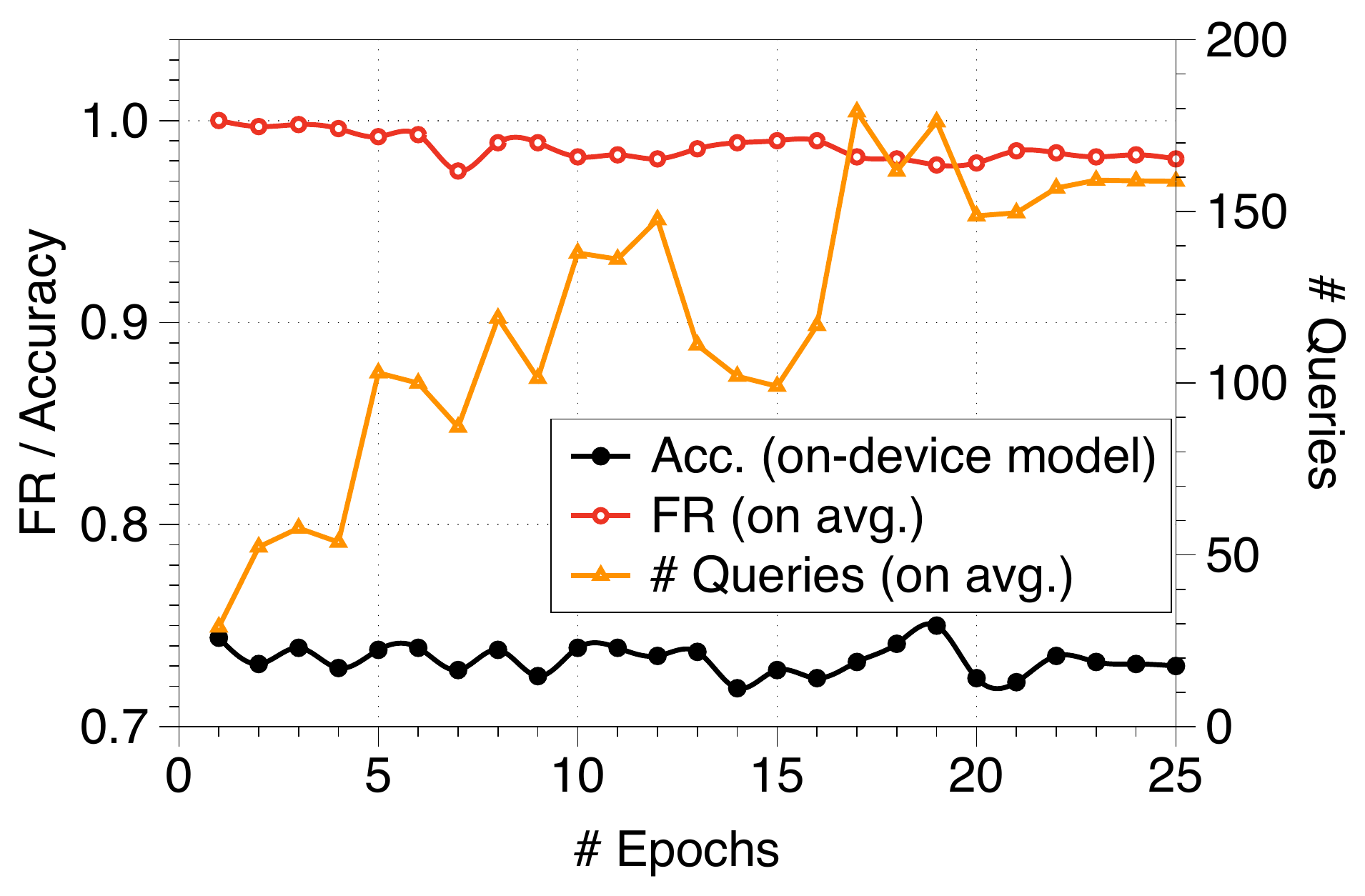}
    \endminipage
    \caption{\textbf{The vulnerability to black-box adversarial attacks while fine-tuning a client-side model.} We run fine-tuning of a ResNet50 in ImageNet for 25 epochs. We show the vulnerability to transfer-based attacks on the left and optimization-based attacks on the right. We find that the vulnerability gradually decreases while preserving the accuracy.}
    \label{fig:finetune-imagenet}
\end{figure}

In \S\ref{sec:defense-evaluations}, 
we observe that the fine-tuning ImageNet models cannot reduce the vulnerability to the same-level as the baselines.
We thus evaluate whether our similarity unpairing objective is ineffective in ImageNet scenarios.
We fine-tune a ResNet50 (ImageNet) model 
with our unpairing objective and measure the model's accuracy 
and the vulnerability to transfer- and optimization-based attacks.
Figure~\ref{fig:finetune-imagenet} shows our results.

We find that the RFR for FGSM decreases from 1.0 to 0.48 at epoch 25.
For PGD-10, the RFR decreases from 1.0 to 0.9.
In the optimization-based attack, 
the query complexity for successful attacks increases from 14 to $\sim$170.
This result indicates that 
our unpairing objective is effective in reducing the vulnerability.
We believe that, by adjusting training hyper-parameters, 
\textit{e.g.}, increasing the learning rate, 
one can suppress the vulnerability similar to the baseline cases.

\section{Evaluating the Power of Robust Models}
\label{appendix:adversarial-training}

Here, we examine whether adversarial training~\cite{PGD, TRADES:ICML2019} (AT) of server-side models,
\textit{i.e.} a standard technique for training robust models,
can provide some benefit for the defender.
There are three ways the defender can employ AT:
(i) adversarially-training only on the server-side models and constructing on-device models from them;
(ii) employing AT only on the on-device models; or
(ii) constructing both the server-side and on-device models. 
We test all three scenarios.

\topic{Setup.}
We conduct the same vulnerability analysis we performed in \S\ref{subsec:characterize-the-vuln}.
\emph{First}, we run AT of ResNet50 to build a robust server-side model.
We use the standard setting,
where we use PGD-7 with the perturbation limit to 8\/255 pixels in $\ell_{\infty}$ norm
and the step-size of 2 pixels.
We then utilize the six mechanisms we previously used 
to construct on-device models in \S\ref{subsec:characterize-the-vuln}.
Note that in the cases where we train a new architecture from scratch, 
\textit{i.e.}, NASNet, MobileNetV2, and SqueezeNet, 
we adversarially-train these models,
which reflects our \emph{second} scenario.
\emph{Third}, we only train robust on-device models 
and use them to attack the undefended server-side model.
%
We perform the transfer-based attacks (FGSM and PGD-10) and 
the query-based attack (P-RGF),
exploiting the on-device models as transfer priors.


\begin{table}[ht]
\centering
\adjustbox{max width=\linewidth}{
    \begin{tabular}{@{}cc|lc|cc|cc@{}}
    \toprule
    \multicolumn{2}{c|}{\textbf{Server $f_s$}} & \multicolumn{2}{c|}{\textbf{Client Models $f_c$}} & \multicolumn{2}{c|}{\textbf{Transfer-}} & \multicolumn{2}{c}{\textbf{Optimization-}} \\ \midrule
    \multicolumn{1}{c}{\textbf{Arch.}} & \multicolumn{1}{c|}{\textbf{Acc.}} & \multicolumn{1}{c}{\textbf{Arch.}} & \textbf{Acc.} & \textbf{FGSM} & \textbf{PGD10} & \textbf{\#Q} & \textbf{FR} \\ \midrule \midrule
    \multirow{7}{*}{\rotatebox[origin=c]{90}{R50}} & \multirow{7}{*}{92\%} & R50 (as-is) & 92\% & 1.00 & 1.00 & 14.1 & 100\% \\
     &  & R50 (ours) & 93\% & 0.06 & 0.08 & 2835.5 & 34\% \\ \cmidrule(l){3-8} 
     &  & R50 & 83\% & 0.03 & 0.27 & 2285.0 & 47\% \\
     &  & R18 & 78\% & 0.04 & 0.30 & 2248.1 & 47\% \\
     &  & NASNet & 75\% & 0.05 & 0.42 & 1787.3 & 59\% \\
     &  & MobileV2 & 82\% & 0.03 & 0.22 & 2429.7 & 44\% \\
     &  & Squeeze & 81\% & 0.03 & 0.25 & 2273.1 & 47\% \\ \bottomrule
    \end{tabular}
}
\caption{\textbf{Impact of robust training on the vulnerability.} We compare our defense with the robust training of on-device models in reducing the vulnerability. The first two rows are the baselines: using the undefended ResNet50 and the model fine-tuned using our defense. The rest are the robust models.}
\label{tbl:robust-models}
\end{table}

\topic{Results.}
Table~\ref{tbl:effectiveness-of-adv-priors} summarizes our results in the first and the second scenario.
As we reviewed in \S\ref{sec:problem}, 
robust models suffer from accuracy degradations---their accuracy is 10--20\% less than the undefended models.
It may not be desirable to sacrifice their server-side model's accuracy.
In transfer-based attacks,
we find that the vulnerability remains the same.
Quantization and pruning lead to the RFR of 0.8--1.0.
In the NASNet, MobileNetV2, and SqueezeNet cases,
we observe the RFR further increases to 0.5--0.6;
they were 0.1--0.2 in our results with non-robust server-side models.
This implies that 
adversarial-training of server-side models 
cannot reduce the vulnerability increase in transfer-based attacks.
Oftentimes, adversarial examples crafted on robust on-device models
transfer better than those crafted on non-robust models.

However, we observe that robust training reduces the vulnerability increase in query-based attacks.
Even in the most vulnerable cases, 
i.e., baseline (as-is), quantization, and pruning,
the FR is 53--64\% and, 
the number of queries needed for crafting a successful adversarial example is 1476--1886.
In other cases, we observe that
the FR is significantly reduced to $\sim$25\%,
and the number of queries the attacker will spend is $\sim$3000.
This implies that in query-based attacks, 
either robust training reduces the vulnerability increase, 
or the attacks are weak 
(i.e., we require future work on stronger query-based attacks to test the vulnerability).

We further show our results, 
reflecting the third scenario, in Table~\ref{tbl:robust-models}.
We first observe that AT is effective in reducing this vulnerability.
Compared to the baseline (as-is), 
AT significantly reduces the RFR of the transfer-based attacks---i.e., from 1.00 to 0.03--0.05 in FGSM and 0.2--0.4 in PGD-10.
In the optimization-based attack, 
the robust models increase the query complexity by two orders of magnitudes and reduce the FR by $\sim$50\%.
However, as expected, the robust on-device models have 10--18\% less accuracy, 
compared to the undefended models we use in \S\ref{sec:exploit-compact-models}.
For comparison, we show that the on-device model fine-tuned with our similarity unpairing (ours) achieves desiderata.
The on-device model has an accuracy of 93\% (no accuracy drop) and reduces the vulnerability more than any robust model.

\section{Evaluating the Power of Robust Architecture}
\label{appendix:robust-architecture}

Here, we additionally examine whether a neural network architecture,
known to be robust to adversarial attacks, can reduce the vulnerability in our settings.

\topic{Setup.}
We use RobNet~\cite{RobNet}, 
a robust architecture found by neural architecture search, for our evaluation.
We first run adversarial-training of RobNet on CIFAR10 
using PGD-7 with the same perturbation bound and step-size we used in our work.
We follow the same training configurations that the original study used.
We then take the pre-trained RobNet (as a server-side model) and
and construct on-device models using efficient deep learning mechanisms.
We test with the most vulnerable cases, i.e., quantization (8-bit) and pruning.
We use those two on-device models as transfer priors
to perform black-box adversarial attacks on the pre-trained RobNet.

\topic{Results.}
We achieve an accuracy of \facc{77}{0.4} on the pre-trained RobNet,
and \facc{77}{0.4} and \facc{75}{0.3} on the quantized and pruned models, respectively.
Our results corroborate the results we had in Appendix~\ref{appendix:adversarial-training}.
We observe that in transfer-based attacks,
the vulnerability stays the same.
The RFR of the FGSM and PGD-10 attacks are 1.00 when we use the 8-bit model.
If we use the sparse model, the RFR is \facc{0.85}{0.09} and \facc{0.84}{0.13} for FGSM and PGD-10.
However, in optimization-based attacks,
we found that the vulnerability increases.
If we use those on-device models, 
the FR (success rate) of P-RGF is 45--55\% 
and the number of queries required to craft an adversarial example is 1547--2186.
The attacker is twice successful at fooling, 
and the query efficiency increases by a factor of 2$\times$,
compared to the setting where we don't have any transfer prior.
We further exploit those on-device models
to perform the query-based attacks on our pre-trained ResNet50.
We observe that the FR and the number of queries required 
are similar to attacking the pre-trained RobNet model.

\section{Details of Adversarial Attacks We Use}
\label{appendix:adversarial-attack-in-details}

We describe the details of the adversarial attacks we use.

\topic{Projected-gradient descent (PGD):}
Given a test-time input $x$, its true label $y$, and a target model $f_{\theta}$, 
PGD~\cite{PGD} works by computing the noise $\delta$ 
that can increase the loss $\mathcal{L}(f_{\theta}(x), y)$.
If the noise is added to the input,
the adversarial example $x'$ = $x + v$ can be misclassified by $f_{\theta}$.
To compute such noise, 
the adversary iteratively computes the gradients
with respect to the input $x$ using $f_{\theta}$.
PGD takes the sign of the gradients
and adjust each element by multiplying the step-size $\alpha$.
PGD bounds the maximum input perturbations (the noise) the attacker can make
by limiting the distance between the input $x$ and the adversarial example $x'$.
Typical choices of the distance is $\ell_{p}$, where $p$=$1, 2,$ or $\infty$,
and, at each iteration, the input gradients are projected to the $\ell_{p}$ space.
When we refer to this attack, 
the number of iterations the adversary will use
is followed by the name PGD, such as PGD-10, for 10 iterations.
If the attacker uses a large bound or more iterations,
the attack becomes stronger; otherwise, it is weak.
We show the PGD algorithm below:
\begin{align*}
    {v}^{t+1} = {\prod}_{|{v}^t|_{p}<\epsilon} { \bigg( v^t + \alpha \text{ } sign \big( \nabla_{v} \mathcal{L}(f_{\theta}(x + v), y) \big) \bigg) },
\end{align*}
where $v$, $t$, $\epsilon$, and $\alpha$ denote the noise, the number of iterations, the bound, and the step-size, respectively.

\topic{Fast-gradient sign method (FGSM):}
FGSM~\cite{FGSM} is another way to craft adversarial examples $v$.
Similar to PGD, 
the attack computes the input gradients $\nabla_{x}\mathcal{L}(f(x), y)$
and add it to the original clean example $x$.
But before the addition, 
the attacker takes only the sign of the input gradient $sign(\cdot)$
and then multiplies it by the step-size $\alpha$.
FGSM leads to adversarial examples, weaker than those from PGD,
but the weaker examples have been used in prior work that studies the transferability
to quantify the model's behaviors nearby its decision boundary.
We show the FGSM algorithm below:
\begin{align*}
    v = {\prod}_{|v|_{p}<\epsilon} { \bigg( \alpha \text{ } sign \big( \nabla_{v} \mathcal{L}(f_{\theta}(x + v), y) \big) \bigg) },
\end{align*}

\topic{Prior-guided random gradient-free (P-RGF):}
Both the PGD and FGSM attacks exploit the transferability, 
\textit{i.e.}, the attacker generates adversarial examples 
on a different model $f_{\theta'}$ and uses them to attack the target model $f_{\theta}$.
In contrast, the P-RGF attack~\cite{P-RGF} 
queries the target model directly 
for crafting adversarial examples.
Black-box attacks typically require
thousands of queries to generate a successful adversarial example.
To reduce such query complexity,
P-RGF exploits a model $f_{\theta'}$ similar to the target
in terms of the training data or model architectures,
as a \emph{transfer-based prior}.
The attack takes advantage of the query information
and the prior for approximating the true input gradients
$\nabla_{x}(f_{\theta}(x), y)$ on $f_{\theta}$.
For the detailed attack algorithms, 
we refer the readers to the original study by Cheng~\textit{et al.}~\cite{P-RGF}.

\end{document}